\documentclass[12pt,draftcls,peerreviewca,onecolumn]{IEEEtran}
\usepackage{multicol}
\usepackage{etoolbox}
\makeatletter
\patchcmd{\@makecaption}
  {\scshape}
  {}
  {}
  {}
\makeatletter
\patchcmd{\@makecaption}
  {\\}
  {.\ }
  {}
  {}
\makeatother

\usepackage{amsfonts}
\usepackage{mathrsfs}
\usepackage{mathtools}
\usepackage{amsfonts}
\usepackage{amssymb}
\usepackage{graphicx}
\usepackage{epsfig}
\usepackage{psfrag}
\usepackage{amsmath}
\usepackage{array}
\usepackage{cases}
\usepackage{eufrak}
\usepackage{cite,graphicx,amsmath,amssymb,color}
\usepackage{algorithmic}
\usepackage{algorithm}
\usepackage{subfigure}
\usepackage{bm}
\usepackage{multirow}
\usepackage{threeparttable}
\usepackage{array}
\usepackage{makecell}

\newtheorem{Thm}{Theorem}
\newtheorem{Lem}{Lemma}

\newtheorem{Prob}{Problem}

\newtheorem{Proof}{Proof}

\newtheorem{rem}{Remark}
\newcommand{\norm}[1]{\left\lVert#1\right\rVert}

\IEEEoverridecommandlockouts

\begin{document}

\setcounter{page}{1}
\title{Multi-Quality Multicast Beamforming based on Scalable Video Coding}
\author{Chengjun Guo,  Ying Cui, Derrick Wing Kwan Ng, and Zhi Liu\thanks{C. Guo and Y. Cui are with Shanghai Jiao Tong University, China. D. W. K. Ng is with the University of New South Wales, Australia. Z. Liu is with Shizuoka  University, Japan. The paper was submitted in part to IEEE GLOBECOM 2017.}}

\maketitle
\thispagestyle{headings}


\begin{abstract}
In this paper, we consider multi-quality multicast beamforming of a video stream from a multi-antenna base station (BS) to multiple  single-antenna users receiving different qualities of the same video stream,  via scalable video coding (SVC). Leveraging the layered structure of SVC and  exploiting superposition coding (SC) as well as  successive interference cancelation (SIC), we propose
a  layer-based multi-quality multicast beamforming scheme. To reduce the complexity, we also propose a  quality-based  multi-quality multicast  beamforming scheme, which further utilizes the layered structure of SVC and quality information of all users.
Under each scheme, for given quality requirements of all users,  we formulate the corresponding optimal beamforming design as a non-convex power minimization problem,
and obtain a globally optimal solution for a class of special cases as well as a locally optimal solution for the general case. Then, we show that the minimum total transmission power of the  quality-based power minimization problem is the same as that of the layer-based power minimization problem,
although the former  incurs a lower computational complexity. Next, we consider the optimal joint layer selection and quality-based multi-quality multicast beamforming design to maximize the total utility representing the satisfaction with the received video quality for all users under a given maximum transmission power budget, which is NP-hard in general. By exploiting the optimal solution of the quality-based power minimization problem, we develop   a greedy algorithm to obtain a near optimal solution. Finally, numerical results show that the proposed solutions achieve better performance than existing solutions.
\end{abstract}


\setcounter{page}{1}
\newpage
\section{Introduction}
Due to the rapidly increasing demands for wireless multimedia applications and the associated energy consumptions, providing higher quality transmissions of videos  over wireless environments at  low energy costs becomes an everlasting endeavor for multimedia service providers \cite{song2008svc,Multiviewvideo2013,MDCvideo2015}. To address this issue, wireless multicast  is considered as a viable solution for delivering   video streams to multiple users simultaneously by effectively utilizing the broadcast nature of the wireless medium.
Wireless multicast has been extensively studied in the scenario where a base station (BS) is equipped with a single antenna \cite{sisocast2005}.
 In practice, deploying multiple antennas at a BS can significantly improve the performance of wireless multicast systems
via designing efficient beamformers.
In \cite{luo2006} and \cite{luo2006-2}, the authors consider multicasting a single message from a multi-antenna BS to a group of single-antenna users, and study the optimal multicast beamforming design to minimize the total transmission power. The design problem is NP-hard in general.
In  \cite{luo2006}, using semidefinite relaxation (SDR) and Gaussian randomization, an approximate  solution is obtained. In \cite{luo2006-2}, by analyzing structural properties of the problem,  a closed-form globally optimal solution is obtained for the two-user case.
As extensions of \cite{luo2006} and \cite{luo2006-2}, the authors in \cite{luo2008} and \cite{luo2008-3} investigate the multi-group multicast case, where multiple independent messages are multicasted simultaneously. Specifically,  in \cite{luo2008}, a suboptimal solution is obtained following a similar approach as in \cite{luo2006}.  In \cite{luo2008-3}, an alternative optimization algorithm is proposed to obtain a near optimal solution. Note that
due to user heterogeneity, the  multicast schemes proposed in
 \cite{luo2006,luo2006-2,luo2008,luo2008-3} usually incur  prohibitively high power costs to guarantee that the user with the worst channel condition in each group can decode the message successfully.


To deal with the problem in video multicast caused by user heterogeneity \cite{luo2006,luo2006-2,luo2008,luo2008-3}, scalable video coding (SVC) has attracted more and more attentions in recent years \cite{schwarz2007overview}. Specifically,
SVC encodes a video stream into one base layer and multiple enhancement layers.
The base layer carries the essential information and provides a minimum quality of the video, and the enhancement layers represent the same video with gradually increasing quality \cite{schwarz2007overview}.
The decoding of a higher enhancement layer is based on the base layer and all its lower enhancement layers.
By multicasting  an SVC-based video,
users with higher channel quality can decode more layers to retrieve the video with  a higher quality  and users with worse channel quality can decode fewer layers to retrieve the video  with a lower quality.


SVC-based video  multicast has application in several scenarios such as  emergency video multicasting  in vehicular ad hoc networks (VANETs), new generation remote teaching  and sports event multicasting \cite{app2011}, where users may have different received video qualities due to the nonuniform cellular usage costs, display resolutions of devices and channel conditions. References \cite{utility07,a0,a2,joint15,layers16,a4,b0} consider SVC-based video multicast and study the optimal layer selection (i.e., select the number of layers  targeted for each user which determines the user's received video quality) and resource allocation (e.g., modulation and coding scheme (MCS) selection \cite{a0,a2,joint15,layers16,a4}, time and frequency resource allocation \cite{layers16,b0}, and power allocation \cite{a4}). They focus on the maximization of the total system utility representing the satisfaction with the received video quality for all users under the total  time \cite{utility07,a0,a2,joint15,layers16,a4}, frequency \cite{layers16} and power \cite{a4} constraints. Due to the discrete nature of the  MCS selection, the optimization problems in \cite{utility07,a0,a2,joint15,layers16,a4} are NP-hard. In \cite{a0} and \cite{a2}, optimal algorithms (with non-polynomial complexity) are developed using dynamic programming.
 In \cite{a4}, the authors  propose an optimal algorithm to solve a simplified problem using dynamic programming. In \cite{utility07,a0,a2,joint15,layers16}, greedy algorithms \cite{utility07,layers16} and heuristic algorithms \cite{joint15,b0} are developed to obtain suboptimal solutions. Note that \cite{utility07,a0,a2,joint15,layers16,a4,b0}  all consider single-antenna BSs and the obtained solutions cannot be directly applied to SVC-based video multicast with multi-antenna BSs.

Recently, the notion of multi-antenna communications in wireless SVC video delivery systems has been pursued.
In \cite{a1,HZN15,choi2015minimum}, the authors consider multi-antenna BSs and study the corresponding SVC-based video multicast schemes.  In particular,  \cite{a1} and \cite{HZN15} study the total utility maximization problem by optimizing the power allocation under a given maximum transmission power budget.
Note that, although adopting multiple antennas at the BS, \cite{a1} and \cite{HZN15} do not consider adaptive beamforming designs for SVC-based video multicast, and hence cannot fully unleash the spatial degrees of freedom in multicasting offered by multiple antennas.
\cite{choi2015minimum} is the first work on the optimal beamforming design for SVC-based video multicast. In particular, \cite{choi2015minimum} employs superposition coding (SC) and successive interference cancellation
(SIC) to improve the performance of SVC-based video multicast.\footnote{SC allows a transmitter to simultaneously send independent messages to multiple receivers. SIC is a decoding technique which decodes multiple signals sequentially by subtracting interference due to the decoded signals before decoding other signals.} However, \cite{choi2015minimum} restricts its study to a special case with two required video qualities targeting for only two users (a near user and a far user).
In addition, the proposed suboptimal algorithm in \cite{choi2015minimum}, which alternatively  optimizes the two beamforming vectors, cannot guarantee the globally minimum transmission power. More importantly, the proposed beamforming design in \cite{choi2015minimum} is not applicable to a general multi-quality multicast scenario with multiple users requiring the same video at different quality levels. 
Therefore, further studies are required to design general multi-quality multicast beamforming for SVC-based video multicast.



In this paper, we consider multi-quality multicast beamforming for SVC-based video multicast from a multi-antenna BS to multiple single-antenna users.  Our  main contributions are summarized below.
\begin{itemize}
\item First, leveraging the layered structure of SVC and  exploiting SC as well as SIC, we propose
a  layer-based multi-quality multicast beamforming scheme to guarantee certain video qualities for all users. When there are multiple consecutive layers which target for the same set of users,  we further propose a quality-based  multi-quality multicast  beamforming scheme, aiming to reduce the complexity.
 \item Then, under each scheme, for given quality requirements of all users, we consider the corresponding optimal multi-quality multicast beamforming design to minimize the total transmission power.
For each optimization problem,  which is non-convex and NP-hard in general,   we obtain a globally optimal solution for a class of special cases using SDR and the rank reduction method, and  a locally optimal solution for the general case using SDR and the penalty method. We also show that the quality-based optimal solution achieves lower total transmission power than  the layer-based optimal solution for some special cases and the same total transmission power as that of the layer-based optimal solution for other cases. In addition, the quality-based optimal solution incurs a much lower computational complexity than the layer-based optimal solution.
\item Next, we consider the optimal joint layer selection and quality-based multi-quality multicast beamforming design to maximize the total utility representing the satisfaction with the received video quality for all users under a given maximum transmission power budget, which is NP-hard in general.
By exploiting the solution of the quality-based power minimization problem  and carefully utilizing SDR, we develop  a greedy algorithm to obtain a near optimal solution.
\item Finally, numerical results show that the proposed solutions provide substantial power saving and total utility improvement compared to existing solutions.
\end{itemize}

\emph{\bf Notation}: Matrices and vectors are denoted by boldfaced uppercase and lowercase characters, respectively.
For a Hermitian matrix $\mathbf{A}$, $\lambda_{\text{max}}(\mathbf{A})$ denotes its maximal eigenvalue and $\mathbf{A}\succeq\mathbf{0}$ means that $\mathbf{A}$ is positive semidefinite. $(\cdot)^{T}$ and $(\cdot)^{H}$ denote the transpose operator and the complex conjugate transpose operator, respectively. $\text{tr}(\cdot)$ and $\text{rank}(\cdot)$ denote the trace and the rank of an input matrix, respectively.
$\text{card}(\cdot)$ denotes the  cardinality of a set. %
$\mathbb{E}[\cdot]$ denotes the statistical expectation. $\mathcal{CN}(\mathbf{a},\mathbf{R})$ represents the distribution of circularly-symmetric complex Gaussian  random vectors with mean vector $\mathbf{a}$ and covariance matrix $\mathbf{R}$. $\mathbf{I}_{N}$ denotes the $N\times N$ identity matrix.  $\mathbb{C}^{N\times M}$ denotes the space of $N\times M$ matrices with complex entries and
$\mathbb{H}^N$ represents the set of all $N\times N$ complex Hermitian matrices. $\succcurlyeq$ indicates element-wise $\geq$.

\section{System Model}

\begin{figure}[t]
\begin{center}
 \includegraphics[width=6in]{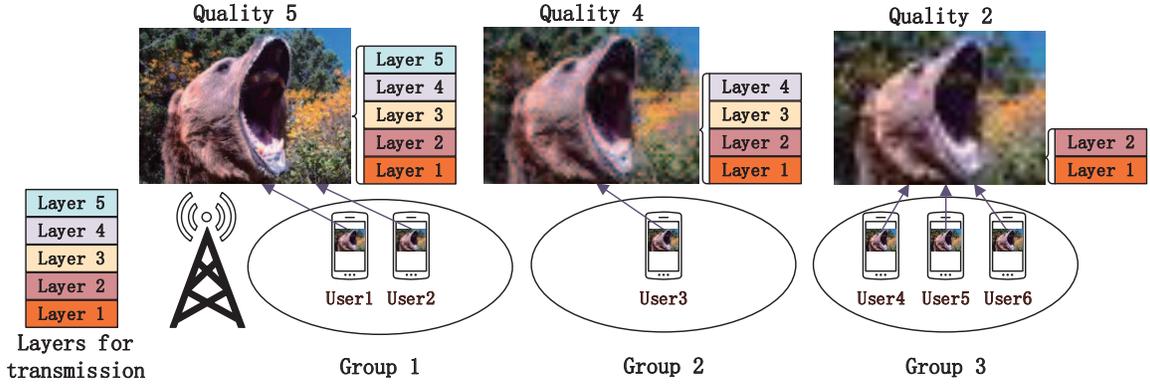}
  \end{center}
  \vspace*{-1cm}
     \caption{\small{The system model of SVC-based video multicast with  $G=3$, $U_1=2$, $U_2=1$, $U_3=3$, $r_{\text{LB},1}=5$, $r_{\text{LB},2}=4$, and $r_{\text{LB},3}=2$.}} 
\label{fig:system-model}
\vspace*{-1cm}
\end{figure}

As illustrated in Fig.~\ref{fig:system-model}, we consider downlink transmissions from a single  base station (BS) to $U(\geq1)$ users. Suppose the locations of all users do not change in the considered timeframe. Denote $\mathcal U:=\{1,2,\cdots, U\}$ as the set of user indices.
The BS is equipped with $N$ transmit antennas and each user is a single-antenna device.
We consider a discrete time system, and assume block fading, i.e., the channel state does not change within one time slot, but changes independently over different time slots.
Let $\mathbf{h}_{u}\in\mathbb{C}^{N\times 1}$ denote the $N\times 1$ complex-valued vector that models the channel from the BS to user $u\in\mathcal U$ at a particular slot.
 We assume that channel state information is available at the BS.

All  users  in the system would like to subscribe a video simultaneously from the BS.  Due to their restricted conditions (such as the cellular usage costs, display resolutions of devices, and channel conditions), users may  require the same video at different quality levels.
Thus, the BS multicasts the video, encoded into $L$ layers using SVC, to all the $U$ users.
For convenience, let $\mathcal{L}\coloneqq \{1,\ldots,L\}$.
According to the encoding mechanism of SVC, the successful decoding of layer $l+1$ requires the successful decoding of layer $l$ and correct receiving of layer $l+1$, for all $l\in\mathcal \{1,\ldots, L-1\}$, and the base layer does not rely on the other layers.
The received video quality can be  improved when more layers are successfully decoded. In particular, $L$ layers correspond to $L$ quality levels, and  layers $1,2,\ldots,l$ yield quality $l$,  for all $l\in\mathcal{L}$.

We consider multiple groups of users.
Later, in Section~\ref{section:power minimization}, we would like to design optimal multi-quality multicast beamforming to minimize the total transmission power under given quality requirements of all $U$ users. We divide the $U$ users into $G$ (requirement-based) groups according to the quality requirements of all $U$ users. That is, users in the same group  have the same quality requirement.
In Section~\ref{section:layer selection}, we would like to optimally assign quality levels to all $U$ users to maximize the total utility of the system under a given maximum transmission
power budget. To reduce the computational complexity, we divide the $U$ users into $G$ (distance-based) groups according to their distances to the BS. That is, users in the same group have similar distances to the BS and the same received video quality.
Let $\mathcal G\coloneqq\{1,2,\cdots, G\}$ denote  the set of group indices. Note that as a special case, there can be exactly one user in each group.
Let $\mathcal{U}_g\subseteq\mathcal{U}$ denote the set of $U_g\coloneqq\text{card}(\mathcal{U}_g)$  users in group $g\in\mathcal G$. Note that $\mathcal U_{g}\cap\mathcal U_{g'}=\emptyset$ for all $g',g\in\mathcal G$, $g'\neq g$, and  $\cup_{g\in\mathcal G}\mathcal U_{g}=\mathcal U$. For convenience, let $\mathbf{U}\coloneqq(U_1,\ldots,U_G)$.


\section{Layer-based and Quality-based Multi-quality Multicast Beamforming Schemes}\label{section:schemes}
In this section, we propose a layer-based multi-quality multicast beamforming scheme and a quality-based multi-quality multicast beamforming scheme. The two schemes adopt multicast beamforming with SC to transmit the SVC-based video from the BS to all the users, and  adopt SIC at each user to facilitate the decoding of its desired layers.
\subsection{Layer-based Multi-quality Multicast Beamforming Scheme}\label{subsec:LB-scheme}

In this part, we propose a layer-based multicast beamforming scheme. First, we introduce some notations.  Let $r_{\text{LB},g}\in\mathcal{L}$ denote the required video quality for the users in group $g\in\mathcal G$.  In other words, the users in group $g\in\mathcal G$ need to decode all the layers in  $\Upsilon_{\text{LB},g}\coloneqq \{1,\ldots,r_{\text{LB},g}\}$ successfully, in order to enjoy the received video at quality  $r_{\text{LB},g}$. For convenience, let $\mathbf{r}_{\text{LB}}\coloneqq(r_{\text{LB},1},\ldots,r_{\text{LB},G})$.
Let $L_{\text{LB},\text{max}}\coloneqq\max\{r_{\text{LB},1},\ldots,r_{\text{LB},G}\}$ denote the total number of layers needed to be transmitted, and let $\Upsilon_{\text{LB},\text{max}}\coloneqq\{1,\ldots,L_{\text{LB},\text{max}}\}$ denote the set of  the layers needed to be transmitted.

We consider layer-based multi-quality multicast beamforming with SC at the BS to transmit all the layers in $\Upsilon_{\text{LB},\text{max}}$.
Let $s_{\text{LB},l}$ represent the signal of layer $l\in\Upsilon_{\text{LB},\text{max}}$. Let $\mathbf{w}_{\text{LB},l}\in\mathbb{C}^{N\times 1}$ denote the $N\times 1$ beamforming vector for signal $s_{\text{LB},l}$. Using layer-based SC, the signal transmitted from the BS to all the users is given by $\sum_{l=1}^{L_{\text{LB},\text{max}}}{\mathbf{w}_{\text{LB},l}s_{\text{LB},l}}$, and the received signal at user $u$ is given by
\begin{align}
y_{\text{LB},u}=\mathbf{h}_u^{H}\left(\sum\limits_{l=1}^{L_{\text{LB},\text{max}}}{\mathbf{w}_{\text{LB},l}s_{\text{LB},l}}\right)+n_{u}, \ u\in\mathcal{U},
\end{align}
where $n_{u}\sim\mathcal{CN}(0,\sigma_{u}^{2})$  represents the noise at user $u\in\mathcal{U}$. Assume $\mathbb{E}[|s_{\text{LB},l}|^2]=1$ for all $l\in\Upsilon_{\text{LB},\text{max}}$ and $\{s_{\text{LB},l}\}_{l=1}^{L_{\text{LB},\text{max}}}$ are mutually uncorrelated. Then, the transmitted power of $s_{\text{LB},l}$ is $\norm{\mathbf{w}_{\text{LB},l}}^2$ and the total transmission power is $\sum_{l=1}^{L_{\text{LB},\text{max}}}{\norm{\mathbf{w}_{\text{LB},l}}^{2}}$. In general, $\norm{\mathbf{w}_{\text{LB},l}}^2>\norm{\mathbf{w}_{\text{LB},l+1}}^2$ for all $l\in\{1,\ldots,L_{\text{LB},\text{max}}-1\}$, as priority is given to lower layers.

We consider layer-based SIC at each user $u\in\mathcal{U}_{g}$ to decode its desired layers in $\Upsilon_{\text{LB},g}$, where $g\in\mathcal{G}$.
 Using SIC, the decoding and cancellation order is always from the stronger received signals to the weaker received signals. Thus, the decoding and cancellation order at user $u\in\mathcal{U}_g$ is $s_{\text{LB},1},\ldots,s_{\text{LB},r_{\text{LB},g}}$. Let $\zeta_{\text{LB},u,l}$ denote the signal-to-interference-plus-noise ratio (SINR) to decode layer $l$ at user $u$ (after removing all the lower layers in $\{1,\ldots,l-1\}$ using SIC when $l\in\{2,\ldots,r_{\text{LB},g}\}$), for all $u\in\mathcal U_g$, $l\in \Upsilon_{\text{LB},g}$ and $g\in\mathcal{G}$.
Thus, we have\footnote{Note that $\sum\limits_{k=L_{\text{LB},\text{max}}+1}^{L_{\text{LB},\text{max}}} {|\mathbf{h}_{u}^{H}\mathbf{w}_{\text{LB},k}|^2}=0$ for notation consistency.}
\begin{align}
\zeta_{\text{LB},u,l}=\frac{|\mathbf{h}_{u}^{H}\mathbf{w}_{\text{LB},l}|^2}{\sum\limits_{k=l+1}^{L_{\text{LB},\text{max}}} {|\mathbf{h}_{u}^{H}\mathbf{w}_{\text{LB},k}|^2+\sigma_{u}^{2}}},\ u\in\mathcal{U}_{g}, l\in\Upsilon_{\text{LB},g}, g\in\mathcal{G}.\label{eq:layer-gamma}
\end{align}
For all $u\in\mathcal{U}_{g}$, $l\in\Upsilon_{\text{LB},g}$ and $g\in\mathcal{G}$, to successfully decode  $s_{\text{LB},l}$ at user $u$, we require $\zeta_{\text{LB},u,l}\geq\Gamma_{\text{LB},l}$, where $\Gamma_{\text{LB},l}\coloneqq2^{R_{\text{LB},l}}-1$ denotes the corresponding SINR threshold for decoding layer $l$ and $R_{\text{LB},l}$ denotes the transmission rate of signal $s_{\text{LB},l}$.
\subsection{Quality-based Multi-quality Multicast Beamforming Scheme}\label{subsec:QB-scheme}
\begin{figure}[t]
\begin{center}
 \includegraphics[width=8cm]{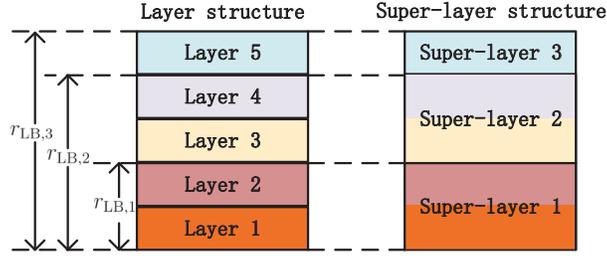}
  \end{center}
  \vspace*{-1cm}
     \caption{\small{The relationship between the layers and the super-layers for the example in Fig.~\ref{fig:system-model}. $G=3$, $r_{\text{LB},1}=5$, $r_{\text{LB},2}=4$, and $r_{\text{LB},3}=2$.}}
\label{fig:relation}
\vspace*{-1cm}
\end{figure}

In this part, we propose a quality-based multi-quality multicast beamforming scheme. In the example shown in Fig.~\ref{fig:system-model}, there are multiple consecutive layers which target for the same user groups. In particular, any user receiving layer 1 also receives layer 2, and any user receiving layer 3 also receives layer 4. Thus, combining layers 1, 2 and layers 3, 4 may potentially reduce the  complexity without sacrificing performance in multi-quality multicast. This motivates us to combine multiple layers into a super-layer based on the quality information of all users, as shown in Fig.~\ref{fig:relation}.

To this end, we construct super-layers from $L_{\text{LB},\text{max}}$ layers in $\Upsilon_{\text{LB},\text{max}}$ based on quality information of all the $G$ groups, as illustrated in Fig.~\ref{fig:relation}.
For ease of illustration, define $r_{\text{LB},G+1}\coloneqq0$ and $\Upsilon_{\text{LB},G+1}\coloneqq\emptyset$.
Let $r_{\text{LB},(1)}\geq r_{\text{LB},(2)}\geq\cdots\geq r_{\text{LB},(G+1)}$ be the arrangement of $r_{\text{LB},g}$ in nondecreasing order. Let $\Upsilon_{\text{LB},(g)}\coloneqq \{1,\ldots,r_{\text{LB},(g)}\}$, for all $g\in\mathcal{G}$.
Obtain the following difference sets: $\Upsilon_{\text{LB},(G)}-\Upsilon_{\text{LB},(G+1)},\ldots,\Upsilon_{\text{LB},(1)}-\Upsilon_{\text{LB},(2)}$.  Let $\Delta\Upsilon_{l}$ denote the $l$-th non-empty difference set in these difference sets. The $l$-th super-layer consists all the layers in $\Delta\Upsilon_{l}$.
Let  $\Upsilon_{\text{QB},\text{max}}\coloneqq\{1, \ldots, L_{\text{QB},\text{max}}\}$ denote the set of all super-layers needed to be transmitted from the BS to all the users in the system, where $L_{\text{QB},\text{max}}$ denotes  the total number of  super-layers needed to be transmitted.
Let $\Upsilon_{\text{QB},g}$ and $r_{\text{QB},g}\coloneqq\text{card}(\Upsilon_{\text{QB},g})$ denote the set of super-layers  and the number of super-layers requested by the users in group $g\in\mathcal{G}$, respectively.

Based on the super-layer structure, we propose a quality-based multi-quality multicast beamforming scheme which resembles the proposed layer-based one in Section~\ref{subsec:LB-scheme}.
Let $\mathbf{w}_{\text{QB},l}\in\mathbb{C}^{N\times 1}$ denote the $N\times 1$ beamforming vector for the signal of super-layer $l$, where $l\in\Upsilon_{\text{QB},\text{max}}$.
The total transmission power is $\sum_{l=l}^{L_{\text{QB},\text{max}}}{\norm{\mathbf{w}_{\text{QB},l}}^{2}}$, and $\norm{\mathbf{w}_{\text{QB},l}}^2>\norm{\mathbf{w}_{\text{QB},l+1}}^2$ in general.
Similarly, the quality-based multi-quality multicast scheme adopts SC at the BS and SIC at each user.
Let $\zeta_{\text{QB},u,l}$ denote the SINR to decode the  signal of super-layer $l$ at user $u$, for all $u\in\mathcal U_g$, $l\in\Upsilon_{\text{QB},g}$ and $g\in\mathcal{G}$.
Thus, we have\footnote{Note that $\sum\limits_{k=L_{\text{QB},\text{max}}+1}^{L_{\text{QB},\text{max}}} {|\mathbf{h}_{u}^{H}\mathbf{w}_{\text{QB},k}|^2}=0$ for notation consistency.}
\begin{align}
\zeta_{\text{QB},u,l}=\frac{|\mathbf{h}_{u}^{H}\mathbf{w}_{\text{QB},l}|^2}{\sum\limits_{k=l+1}^{L_{\text{QB},\text{max}}} {|\mathbf{h}_{u}^{H}\mathbf{w}_{\text{QB},k}|^2+\sigma_{u}^{2}}},\ u\in\mathcal{U}_{g},l\in\Upsilon_{\text{QB},g}, g\in\mathcal{G}.\label{eq:quality-gamma}
\end{align}
For all $u\in\mathcal{U}_{g}$, $l\in\Upsilon_{\text{QB},g}$ and $g\in\mathcal{G}$, to successfully decode the signal of super-layer $l$ at user $u$, we require $\zeta_{\text{QB},u,l}\geq\Gamma_{\text{QB},l}$, where $\Gamma_{\text{QB},l}\coloneqq2^{R_{\text{QB},l}}-1$ denotes the corresponding SINR threshold for super-layer $l$ and $R_{\text{QB},l}\coloneqq \sum_{j\in\Delta\Upsilon_l}R_{\text{LB},j}$ denotes the transmission rate of the signal of super-layer $l$.

\section{Total Transmission Power Minimization}\label{section:power minimization}
In this section, we consider the power-efficient multi-quality multicast beamforming design in one time slot, the duration of which is about $1-5$ milliseconds in practical systems, e.g., LTE, under given quality requirements of all users $\mathbf{r}_{\text{LB}}$.  Under each scheme proposed in Section~\ref{section:schemes},  we formulate the corresponding optimal beamforming design as a
non-convex power minimization problem, and obtain a globally optimal solution for a class of special
cases as well as a locally optimal solution for the general case. Then, we compare the layer-based and quality-based  optimal solutions in terms of the total transmission power and computational complexity. As $\mathbf{r}_{\text{LB}}$ is given, in this section, without loss of generality (w.l.o.g.), we assume $r_{\text{LB},1}>r_{\text{LB},2}>\cdots>r_{\text{LB},G}$. Thus, we have $L_{\text{LB},\text{max}}=r_{\text{LB},1}\geq G$. Besides, according to the construction method of super-layers, we have   $r_{\text{QB},g}=G+1-g$ for all $g\in\mathcal{G}$, and $L_{\text{QB,\text{max}}}=r_{\text{QB},1}=G$.
\subsection{Problem Formulation}
For given quality requirements of all users $\mathbf{r}_{\text{LB}}$, we would like to design the optimal layer-based (quality-based) beamformer to minimize the total transmission power for the layer-based (quality-based) multi-quality multicast beamforming scheme under the successful decoding constraints. As the two schemes share similar expressions for the total transmission power and SINRs, we have the following unified formulation.
\begin{Prob} [Power Minimization]\label{P1}
For all $i\in\{\text{LB},\text{QB}\}$,
\begin{align}
 P^{\star}_{i}\triangleq\min_{\{\mathbf{w}_{i,l}\in\mathbb{C}^{N\times1}\}_{l=1}^{L_{i,\text{max}}}} &\sum_{l=1}^{L_{i,\text{max}}}\norm{\mathbf{w}_{i,l}}^2\\
    \mathrm{s.t.} ~~\quad  &\frac{|\mathbf{h}_{u}^{H}\mathbf{w}_{i,l}|^2}{\sum\limits_{k=l+1}^{L_{i,\text{max}}} {|\mathbf{h}_{u}^{H}\mathbf{w}_{i,k}|^2+\sigma_{u}^{2}}}\geq \Gamma_{i,l}, \ u\in\mathcal{U}_{g},l\in\Upsilon_{i,g},g\in\mathcal{G},\label{constr:P1-sinr}
\end{align}
where $P^{\star}_{i}$ denotes  the  optimal value of the above optimization problem.\nonumber
\end{Prob}

Note that Problem~\ref{P1} with $i=\text{LB}$ indicates the power minimization problem for the layer-based multi-quality multicast beamforming scheme, and Problem~\ref{P1} with $i=\text{QB}$ represents the power minimization problem for the quality-based multi-quality multicast beamforming scheme.

\begin{rem}[Illustration of Problem~\ref{P1} for Layer-based Scheme]\label{rem:SCP1}
Problem~\ref{P1} for the layer-based scheme is a generalization of the traditional multicast beamforming problem \cite{luo2006}  (i.e., multicasting a common message to multiple users) and the recently studied two-quality multicast beamforming  problem \cite{choi2015minimum} (i.e.,  multicasting a video with two qualities to two users). In particular, Problem~\ref{P1} for the layer-based scheme with $G=1$ and $L_{\text{LB},\text{max}}=1$ is identical to the  traditional one in \cite{luo2006},
and Problem~\ref{P1} for the layer-based scheme with $G=2$, $U_1=1$, $U_2=1$ and $L_{\text{LB},\text{max}}=2$ is the same as the one in \cite{choi2015minimum}. Note that the methods in \cite{luo2006,choi2015minimum} cannot be applied to the case considered in this paper directly.
\end{rem}

\subsection{Optimal Solution}\label{subsec:power-optimal}
To solve Problem~\ref{P1}, we first analyze its structure. Considering any $i\in\{\text{LB},\text{QB}\}$,
Problem~\ref{P1} is a quadratically constrained quadratic programming (QCQP) problem. The objective function is convex while the constraints are non-convex.  Problem~\ref{P1} is NP-hard in general \cite{luo2006}. To obtain a globally optimal solution, an exhaustive search method  is required, which incurs  a prohibitively high computational complexity. This motivates the pursuit of approximate solutions of Problem~\ref{P1}, which have low complexity and promising performance. Towards this end, we first introduce an auxiliary optimization matrix $\mathbf{X}_{i,l}\coloneqq \mathbf{w}_{i,l}\mathbf{w}_{i,l}^{H}\in\mathbb{H}^N$, for all $l\in\Upsilon_{i,\text{max}}$. Note that $\mathbf{X}_{i,l}=\mathbf{w}_{i,l}\mathbf{w}_{i,l}^{H}\in\mathbb{H}^{N}$ for some $\mathbf{w}_{i,l}\in\mathbb{C}^{N}$ if and only if $\mathbf{X}_{i,l}\succeq\mathbf{0}$ and $\text{rank}(\mathbf{X}_{i,l})=1$ \cite{luo2006}.
Defining  $\mathbf{C}_{u}\coloneqq \mathbf{h}_{u}\mathbf{h}_{u}^{H}$ for all $u\in\mathcal{U}$, Problem~\ref{P1} can be equivalently reformulated as follows.

\begin{Prob} [Equivalent Problem of Problem~\ref{P1}]\label{EFP1}
For all $i\in\{\text{LB},\text{QB}\}$,
\begin{align}
 P^{\star}_{i}=\min_{\{\mathbf{X}_{i,l}\in\mathbb{H}^N\}_{l=1}^{L_{i,\text{max}}}} &\sum_{l=1}^{L_{i,\text{max}}}\text{tr}(\mathbf{X}_{i,l})\label{objf:EP1}\\
    \mathrm{s.t.} ~\quad  &\text{tr}(\mathbf{C}_{u}\mathbf{X}_{i,l})\geq\Gamma_{i,l}\left(\sum\limits_{k=l+1}^{L_{i,\text{max}}} {\text{tr}(\mathbf{C}_{u}\mathbf{X}_{i,k})+\sigma_{u}^{2}}\right),~u\in\mathcal{U}_{g},l\in\Upsilon_{i,g}, g\in\mathcal{G},\label{constr:EP1-sinr}\\
  &\mathbf{X}_{i,l}\succeq\mathbf{0},\  l\in\Upsilon_{i,\text{max}},\label{constr:EP1-sd}\\
    &\text{rank}(\mathbf{X}_{i,l})=1, \ l\in\Upsilon_{i,\text{max}}.\label{constr:EP1-rank}
\end{align}
\end{Prob}

In Problem~\ref{EFP1}, the objective function and the constraints  in \eqref{constr:EP1-sinr} and \eqref{constr:EP1-sd} are  affine with respect to $\mathbf{X}_{i,l}$,  while the rank-one constraints in \eqref{constr:EP1-rank} are non-convex. Thus,  Problem~\ref{EFP1} is non-convex. By dropping the rank-one constraints in \eqref{constr:EP1-rank}, we can obtain the following SDR \cite{JR:Layered_SDP} of Problem~\ref{P1}.

\begin{Prob} [SDR of Problem \ref{P1}]\label{SDR1}
For all $i\in\{\text{LB},\text{QB}\}$,
\begin{align}
  P_{i,\text{SDR}}^{\star}\triangleq\min_{\{\mathbf{X}_{i,l}\in\mathbb{H}^{N}\}_{l=1}^{L_{i,\text{max}}}} &\sum_{l=1}^{L_{i,\text{max}}}\text{tr}(\mathbf{X}_{i,l})\\
    \mathrm{s.t.} ~\quad &\eqref{constr:EP1-sinr},~\eqref{constr:EP1-sd},\nonumber
\end{align}
  where $P_{i,\text{SDR}}^{\star}$ denotes the optimal value of the above optimization problem.\nonumber
\end{Prob}

Problem~\ref{SDR1} is a standard semidefinite programming (SDP) problem which  is convex and can be efficiently solved by modern SDP solvers. It is obvious that $P_{i,\text{SDR}}^{\star}\leq P^{\star}_{i}$ for all $i\in\{\text{LB},\text{QB}\}$. Thus, we refer to $P_{\text{LB},\text{SDR}}^{\star}$ and $P_{\text{QB},\text{SDR}}^{\star}$ as the layer-based SDR lower bound and the quality-based SDR lower bound, respectively.
If Problem~\ref{SDR1} has a rank-one optimal solution, i.e., $\mathbf{X}_{i,l}=\mathbf{w}_{i,l}\mathbf{w}_{i,l}^H\in\mathbb{H}^N$ for some $\mathbf{w}_{i,l}\in\mathbb C^{N\times 1}$, for all $l\in\Upsilon_{i,\text{max}}$,  then $\{\mathbf{w}_{i,l}\}_{l=1}^{L_{i,\text{max}}}$ is also an optimal solution of Problem~\ref{P1}, implying $P_{i,\text{SDR}}^{\star}=P^{\star}_{i}$.

Next, we obtain a globally optimal solution of Problem~\ref{P1} for a class of special cases as well as a locally optimal solution for the general case.
\subsubsection{Special Case}\label{subsubsec:special-layer}

\begin{table}[t]
\caption{The class of special cases of problem~\ref{P1} for the layer-based scheme.}\label{table1}
\vspace{-0.9cm}
\scriptsize
\begin{center}
\begin{tabular}{|l|l|l|}
\hline
\makecell[c]{$G$} &\makecell[c]{$\mathbf{U}$} &\makecell[c]{$\mathbf{r}_{\text{LB}}$}\\
\hline
\multirow{3}{*}{2}&\multirow{2}{*}{$U_1=1,U_2=1$}&$r_{\text{LB},1}\in\{2,\ldots,L\},r_{\text{LB},2}=1$\\
\cline{3-3}
&\multirow{2}{*}{}&$r_{\text{LB},1}\in\{3,\ldots,L\},r_{\text{LB},2}=2$\\
\cline{2-3}
&$U_1=1,U_2=2$ &$r_{\text{LB},1}\in\{2,\ldots,L\},r_{\text{LB},2}=1$\\
\hline
\multirow{3}{*}{1}&$U_1=1$&$r_{\text{LB},1}\in\mathcal{L}$\\
\cline{2-3}
&$U_1=2$&$r_{\text{LB},1}\in\{1,2\}$\\
\cline{2-3}
&$U_1=3$&$r_{\text{LB},1}=1$\\
\hline
\end{tabular}
\end{center}
\vspace{-0.5cm}
\end{table}

\begin{table}[t]
\caption{The class of  special cases of  problem~\ref{P1} for the quality-based scheme.}\label{quality-based special case}
\vspace{-0.9cm}
\scriptsize
\begin{center}
\begin{tabular}{|l|l|l|}
\hline
\makecell[c]{$G$} &\makecell[c]{$\mathbf{U}$} &\makecell[c]{$\mathbf{r}_{\text{LB}}$}\\
\hline
\multirow{2}{*}{2}&$U_1=1,U_2=1$&$r_{\text{LB},1}={\iota}',r_{\text{LB},2}={\iota}$, where ${\iota}'>{\iota}$, and ${\iota}',{\iota}\in\mathcal{L}$\\
\cline{2-3}
&$U_1=1,U_2=2$ &$r_{\text{LB},1}={\iota}',r_{\text{LB},2}={\iota}$, where ${\iota}'>{\iota}$, and ${\iota}',{\iota}\in\mathcal{L}$\\
\hline
1&$U_1\in\{1,2,3\}$&$r_{\text{LB},1}\in\mathcal{L}$\\
\hline
\end{tabular}
\end{center}
\vspace{-1.3cm}
\end{table}

First, for all $i\in\{\text{LB},\text{QB}\}$,  we study the tightness of the adopted SDR.  Consider the class of special cases of Problem~\ref{P1}, in which the system parameters $(G,\mathbf{U},\mathbf{r}_{\text{LB}})$ satisfy:
\begin{align}\label{rank-reduction-constnum3}
\sum_{g=1}^{G}U_gr_{i,g}\leq L_{i,\text{max}}+2,
\end{align}
where $\sum_{g=1}^{G}U_{g}r_{i,g}$ represents the total number of constraints of Problem~\ref{SDR1}.
In particular, the class of special cases of Problem~\ref{P1} with $i=\text{LB}$ is denoted by $\mathcal{S}_{\text{LB},\text{opt}}\coloneqq\{(G,\mathbf{U},\mathbf{r}_{\text{LB}})|~\eqref{rank-reduction-constnum3} ~\text{is satisfied for}~i=\text{LB}\}$ and is  illustrated in Table~\ref{table1}.
Note that \cite{choi2015minimum} considers the case of $G=2,U_1=U_2=1,r_{\text{LB},1}=2$ and $r_{\text{LB},2}=1$, which belongs to $\mathcal{S}_{\text{LB},\text{opt}}$, and  proposes an iterative algorithm
to obtain a locally optimal solution (which cannot guarantee the globally minimum transmission power), by optimizing $\mathbf{w}_{\text{LB},1}$ and $\mathbf{w}_{\text{LB},2}$ alternatively.
In addition, the class of special cases of Problem~\ref{P1} with $i=\text{QB}$ is denoted by   $\mathcal{S}_{\text{QB},\text{opt}}\coloneqq\{(G,\mathbf{U},\mathbf{r}_{\text{LB}})|~\eqref{rank-reduction-constnum3}~\text{is satisfied for}~i=\text{QB}\}$ and is illustrated in Table~\ref{quality-based special case}.
As one $\mathbf{r}_{\text{QB}}\coloneqq(r_{\text{QB},1},\ldots,r_{\text{QB},2})$ corresponds to multiple $\mathbf{r}_{\text{LB}}$,  we have  $\mathcal{S}_{\text{QB},\text{opt}}\supseteq\mathcal{S}_{\text{LB},\text{opt}}$, which can be easily seen by comparing Table~\ref{table1} and Table~\ref{quality-based special case}.

Next, for all $i\in\{\text{LB},\text{QB}\}$, we propose an algorithm to obtain a globally optimal solution of Problem~\ref{P1} for the class of  special cases in $\mathcal{S}_{i,\text{opt}}$, by constructing a rank-one optimal solution of Problem~\ref{SDR1} using SDP and the rank reduction method in \cite{huang2010rank}.
First, using the arguments in  \cite{huang2010rank}, we show that  for the considered class of special cases in $\mathcal{S}_{i,\text{opt}}$, Problem~\ref{SDR1}  has a rank-one optimal solution, which is also optimal to Problem~\ref{P1}, i.e., $P_{i,\text{SDR}}^{\star}=P^{\star}_{i}$.
By \cite{huang2010rank}, we know that there exists an optimal solution $\{\mathbf{X}_{i,l}^{\star}\}_{l=1}^{L_{i,\text{max}}}$ of  Problem~\ref{SDR1}   satisfying $\sum_{l=1}^{L_{i,\text{max}}}\text{rank}^{2}(\mathbf{X}_{i,l}^{\star})\leq \sum_{g=1}^{G}U_{g}r_{i,g}.$
 Thus, by \eqref{rank-reduction-constnum3},
 we have
 \begin{align}
\sum_{l=1}^{L_{i,\text{max}}}\text{rank}^{2}(\mathbf{X}_{i,l}^{\star})\leq L_{i,\text{max}}+2.\label{rank-reduction-constnum2}
 \end{align}
Since  the rank of a matrix is an integer and \eqref{constr:EP1-sinr} forces that zero matrices cannot be a feasible solution for $\zeta_{i,u,l}>0$,  \eqref{rank-reduction-constnum2} is equivalent to $\text{rank}(\mathbf{X}_{i,l}^{\star})=1$,  for all $l\in\Upsilon_{i,\text{max}}$. Therefore, we can show that Problem~\ref{SDR1} has a rank-one optimal solution for the class of special cases in $\mathcal{S}_{i,\text{opt}}$.
Next, we find a rank-one optimal solution of Problem~\ref{SDR1} for this class of special cases. Specifically, we first obtain an optimal  solution (with arbitrary ranks) of Problem~\ref{SDR1} using SDP. Then, we apply the rank reduction method proposed in \cite{huang2010rank} to obtain a rank-one optimal solution of Problem~\ref{SDR1} by gradually reducing the rank of the optimal solution obtained in the first step.  Algorithm~\ref{alg:rank-reduction} summarizes the procedures to obtain an optimal solution $\{\mathbf{w}^{\star}_{i,l}\}_{l=1}^{L_{i,\text{max}}}$ of Problem~\ref{P1}.

\begin{figure}[t]
\begin{center}
 \subfigure[\small{Layer-based optimal solution. $r_{\text{LB},1}=2, r_{\text{LB},2}=1.$}]
 {\resizebox{5.4cm}{!}{\includegraphics{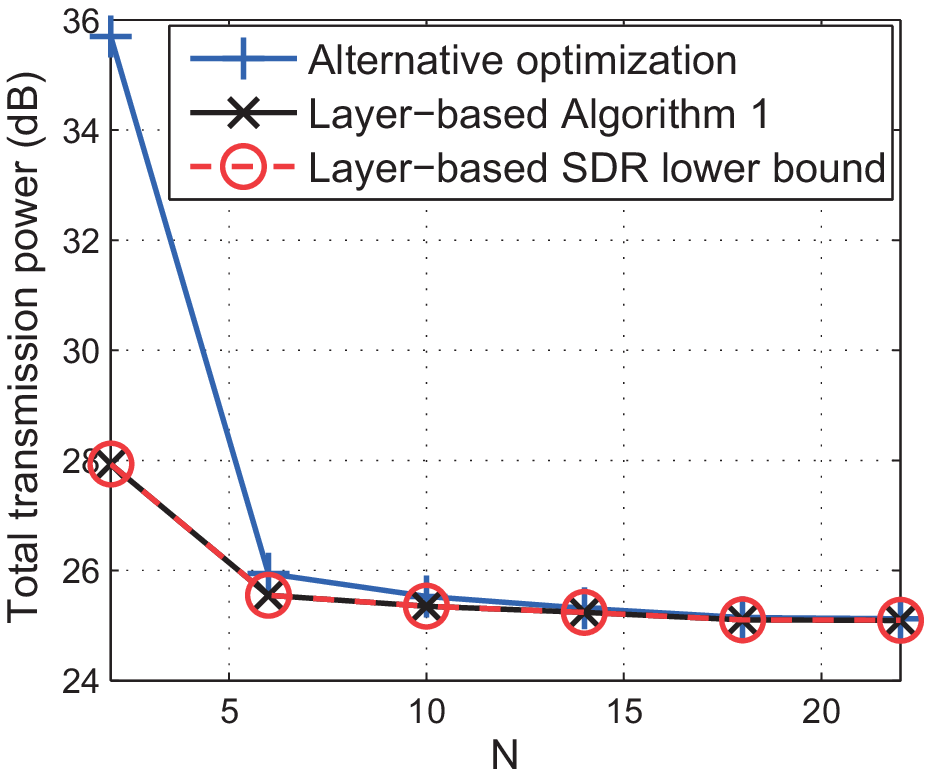}}}
 \subfigure[\small{Quality-based optimal solution. $r_{\text{LB},1}=3, r_{\text{LB},2}=2, R_{\text{LB},3}=2.$}]
 {\resizebox{5.4cm}{!}{\includegraphics{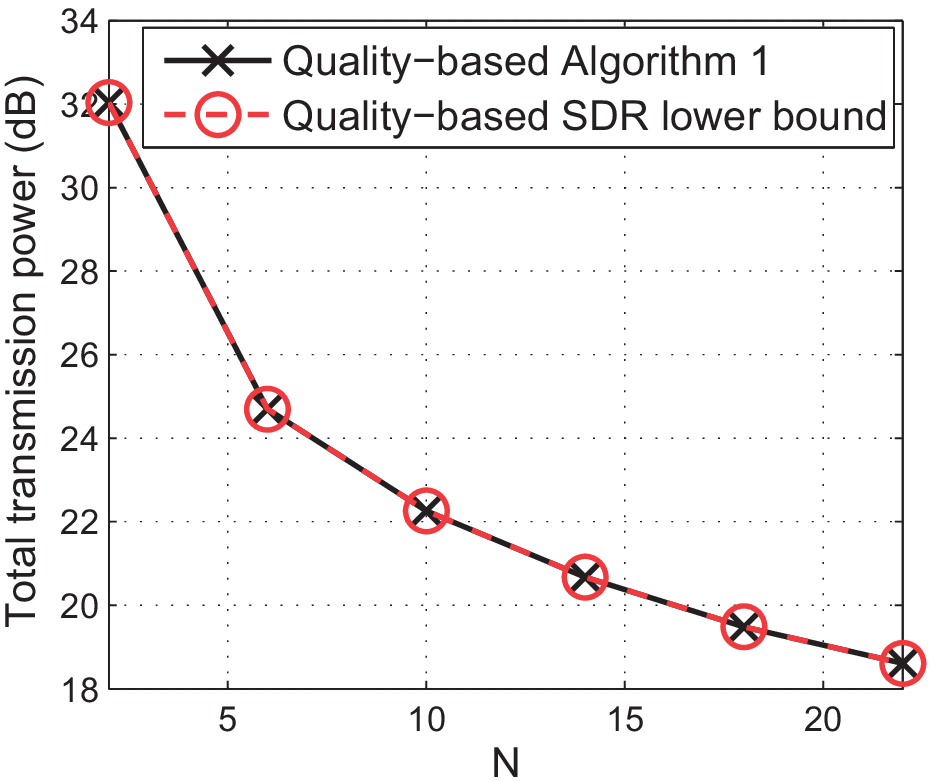}}}
 \end{center}
 \vspace*{-0.3cm}
   \caption{\small{Total transmission power versus the number of antennas at the BS.~Assume that the users in the same group have the same distance to BS. We denote $d_u$ as the relative distance between the BS and user $u$ for all $u\in\mathcal{U}$ (provided that the distance between the BS and the users in group $1$ is normalized) and denote $\eta$ as the path loss exponent. $G=2$, $U_1=U_2=1$, $R_{\text{LB},1}=R_{\text{LB},2}=2$,  $\sigma_{1}^{2}=\sigma_{2}^{2}=1$, $\mathbf{h}_{1}\sim\mathcal{CN}(0,\frac{1}{d_1^{\eta}}\mathbf{I}_{N})$, $\mathbf{h}_{2}\sim\mathcal{CN}(0,\frac{1}{d_2^{\eta}}\mathbf{I}_{N})$, $\eta=2$, $d_1=1$ and $d_2=2$ \cite{choi2015minimum}. The total transmission power presented in this paper is obtained by averaging over 1000 randomly chosen global channel states.}}
   \label{fig:simulation-sc}
   \vspace{-0.2cm}
\end{figure}

Finally, we demonstrate the optimality of  Algorithm~\ref{alg:rank-reduction} using numerical results.
In the simulation, we consider a setup similar to that in  \cite{choi2015minimum}. For $i=\text{LB}$, from Fig.~\ref{fig:simulation-sc} (a), we observe that the performance of Algorithm~\ref{alg:rank-reduction} and the layer-based SDR lower bound of Problem~\ref{SDR1} obtained using SDP are identical, verifying the optimality of Algorithm~\ref{alg:rank-reduction}. In addition,
the minimum  transmission power achieved by Algorithm~\ref{alg:rank-reduction} is always less than or equal to that obtained using the alternative optimization method in \cite{choi2015minimum}. The performance gain of the proposed Algorithm~\ref{alg:rank-reduction} over the alternative optimization method increases, as the number of antennas at the BS decreases.
For $i=\text{QB}$,
from Fig.~\ref{fig:simulation-sc} (b), we can observe that the performance of Algorithm~\ref{alg:rank-reduction} and the optimal value of Problem~\ref{SDR1} obtained using SDP are always the same.

\begin{algorithm}\label{Rank Reduction for Problem 2}
    \caption{\small{Globally Optimal Solution of Problem~\ref{P1} for Special Cases}}
\begin{multicols}{2}
\begin{footnotesize}
     \begin{algorithmic}[1]
           \STATE   Find an optimal solution $\{\mathbf{X}_{i,l}\}_{l=1}^{L_{i,\text{max}}}$ (with arbitrary ranks) of Problem~\ref{SDR1}; \\
           \STATE  Evaluate $\gamma_{l}=\text{rank}(\mathbf{X}^{\star}_{i,l})$, for all $l\in\Upsilon_{i,\text{max}}$, and set $\gamma=\sum\limits_{l=1}^{L_{i,\text{max}}}\gamma_{l}^{2}$;\\
        \WHILE{$\gamma>L_{i,\text{max}}+2$}
      \STATE    For all $l\in\Upsilon_{i,\text{max}}$, decompose $\mathbf{X}_{i,l}^{\star}=\mathbf{V}_{l}\mathbf{V}_{l}^H$;\\
      \STATE    Find a nonzero solution $\{\mathbf{\Delta}_{l}\}_{l=1}^{L_{i,\text{max}}}$ of the system of linear equations:\\
      $\text{tr}(\mathbf{V}_{l}^{H}\mathbf{C}_{u}\mathbf{V}_{l}\mathbf{\Delta}_{l})-\Gamma_{l}\sum\limits_{k=l+1}^{L_{i,\text{max}}}\text{tr}(\mathbf{V}_{k}^{H}\mathbf{C}_{u}\mathbf{V}_{k}\mathbf{\Delta}_{k})=0$, $u\in\mathcal{U}_g, l\in\Upsilon_{i,g}, g\in\mathcal{G}$,\\
      where $\mathbf{\mathbf{\Delta}}_{l}$ is a $\gamma_{l}\times\gamma_{l}$ Hermitian matrix for all $l=1,\ldots,L_{i,\text{max}}$;
      \STATE    Evaluate the eigenvalues $\delta_{l1},\ldots,\delta_{l\gamma_{l}}$ of $\mathbf{\Delta}_{l}$ for all $l=1,\ldots,L_{i,\text{max}}$;\\
      \STATE    Determine $l_{0}$ and $k_{0}$ such that\\
      $|\delta_{l_{0}k_{0}}|=\text{max}\{|\delta_{lk}|:1\leq k\leq\gamma_{l}, l\in\Upsilon_{i,\text{max}}\}$;\\
      \STATE    Compute $\mathbf{X}_{i,l}^{\star}=\mathbf{V}_{l}(\mathbf{I}_{\gamma_{l}}-(1/\delta_{l_{0}k_{0}})\mathbf{\Delta}_{l})\mathbf{V}_{l}^H$, for all $l=1,\ldots,L_{i,\text{max}}$;\\
      \STATE    Evaluate $\gamma_{l}=\text{rank}(\mathbf{X}_{i,l}^{\star})$, for all $l\in\Upsilon_{i,\text{max}}$, and set $\gamma=\sum\limits_{l=1}^{L_{i,\text{max}}}\gamma_{l}^{2}$;\\
      \ENDWHILE
      \STATE   Decompose $\mathbf{X}_{i,l}^{\star}=\mathbf{w}_{i,l}^{\star}(\mathbf{w}_{i,l}^{\star})^H$, for all $l=1,\ldots,L_{i,\text{max}}$.
    \end{algorithmic}
    \end{footnotesize}\label{alg:rank-reduction}
    \end{multicols}
\end{algorithm}

\subsubsection{General Case}
For all $i\in\{\text{LB},\text{QB}\}$, for the general case (not in $\mathcal{S}_{i,\text{opt}}$),  a rank-one optimal solution of Problem~\ref{SDR1} cannot always be obtained by Algorithm~\ref{alg:rank-reduction},  as \eqref{rank-reduction-constnum2} does not always hold.  In this case, we apply the penalty method proposed in \cite{phan2012nonsmooth}
to obtain a locally optimal solution of Problem~\ref{P1}. In particular, as in \cite{phan2012nonsmooth}, we first  transform the  rank-one constraints in \eqref{constr:EP1-rank}  to the following equivalent reverse convex constraint \cite{phan2012nonsmooth}:
\begin{align}
\sum_{l=1}^{L_{i,\text{max}}}\left(\text{tr}(\mathbf{X}_{i,l})-\lambda_{\text{max}}(\mathbf{X}_{i,l})\right)\leq0.\label{eq:reverse convex}
\end{align}
As a result, we can consider an equivalent optimization problem for Problem~\ref{EFP1}  which minimizes the objective function in \eqref{objf:EP1} under the constraints in \eqref{constr:EP1-sinr}, \eqref{constr:EP1-sd} and \eqref{eq:reverse convex}. Note that this new optimization problem is also non-convex due to the reverse convex constraint  in~\eqref{eq:reverse convex}.
As in \cite{phan2012nonsmooth}, we augment the objective function by introducing a concave penalty function utilizing the left hand side of \eqref{eq:reverse convex}. Then, the resulting transformed problem is handled by  using an iterative convex-concave method.
Algorithm~\ref{alg:penalty method} summarizes the procedures to produce a rank-one locally optimal solution $\{\mathbf{X}_{i,l}^{\dagger}\}_{l=1}^{L_{i,\text{max}}}$ of Problem~\ref{SDR1},
which corresponds to an optimal solution of  Problem~\ref{P1}~\cite{phan2012nonsmooth}.
\begin{algorithm}
    \caption{Locally Optimal Solution of Problem~\ref{P1} for the General Case}
\begin{multicols}{2}
\begin{footnotesize}
        \begin{algorithmic}[1]
           \STATE \textbf{Initial Step}: Initialize penalty factor $\mu$ and obtain an optimal solution, $\{\mathbf{X}_{i,l}^{(0)}\}_{l=1}^{L_{i,\text{max}}}$ to Problem~\ref{SDR1} using SDP. Set $\kappa\coloneqq0$.\\

           \STATE \textbf{Step $\kappa$}: Find an optimal solution $\{\mathbf{X}_{i,l}^{(\kappa+1)}\}_{l=1}^{L_{i,\text{max}}}$ to minimize $\sum_{l=1}^{L_{i,\text{max}}}(\text{tr}(\mathbf{X}_l)+\mu(\text{tr}(\mathbf{X}_{i,l})-\lambda_{\text{max}}(\mathbf{X}_{i,l})-\text{tr}(\mathbf{x}_{i,l}^{(\kappa)}\mathbf{x}_{i,l}^{(\kappa)H}, \mathbf{X}_{i,l}-\mathbf{X}_{i,l}^{(\kappa)})))$ subject to \eqref{constr:EP1-sinr} and \eqref{constr:EP1-sd}, where $\mathbf{x}_{i,l}$ is the unit-norm eigenvector of $\mathbf{X}_{i,l}$ corresponding to $\lambda_{\text{max}}(\mathbf{X}_{i,l})$.\\
           \IF{$\text{tr}(\mathbf{X}_{i,l}^{(\kappa+1)})\approx\lambda_{\text{max}}(\mathbf{X}_{i,l}^{(\kappa+1)})$ and $\text{tr}(\mathbf{X}_{i,l}^{(\kappa+1)})\approx\text{tr}(\mathbf{X}_{i,l}^{(\kappa)})$ for all $l\in\Upsilon_{i,\text{max}}$}
           \STATE Set $\mathbf{X}_{i,l}^{\dagger}\coloneqq\mathbf{X}_{i,l}$ for all $l\in\Upsilon_{i,\text{max}}$. Terminate and output $\{\mathbf{X}_{i,l}^{\dagger}\}_{l=1}^{L_{i,\text{max}}}$.\\
           \ELSIF{$\mathbf{X}_{i,l}^{(\kappa+1)}\approx\mathbf{X}_{i,l}^{(\kappa)}$ for all $l\in\Upsilon_{i,\text{max}}$ }
           \STATE Reset $\mu\coloneqq 2\mu$ and return to Initial Step.\\
           \ELSE
           \STATE Reset $\kappa\coloneqq\kappa+1$, $\mathbf{X}_{i,l}^{(\kappa)}\coloneqq\mathbf{X}_{i,l}^{(\kappa+1)}$  for all $l\in\Upsilon_{i,\text{max}}$  {and return to Step $\kappa$}.\\
           \ENDIF
    \end{algorithmic}
    \end{footnotesize}\label{alg:penalty method}
    \end{multicols}
    \vspace*{-0.2cm}
\end{algorithm}
\subsection{Comparison of Layer-based and Quality-based Optimal Solutions}
In this part, we  compare the  layer-based and quality-based optimal solutions in terms of the total transmission power and computational complexity, for the same system parameters. Note that when $L_{\text{LB},\text{max}}=G$ (i.e., $r_{\text{LB},1}=G, r_{\text{LB},2}=G-1,\ldots,r_{\text{LB},G}=1$), Problem~\ref{P1} with $i=\text{LB}$ and Problem~\ref{P1} with $i=\text{QB}$ share exactly the same formulation, and hence the same optimal value and computational complexity. Thus, in the following comparison, we focus on the case where $L_{\text{LB},\text{max}}>G$.

\subsubsection{Total Transmission Power Analysis}
To analyze the relationship between the minimum  total transmission powers of Problem~\ref{P1} with $i=\text{LB}$ and $i=\text{QB}$, we first explore optimality properties of Problem~\ref{P1}.
\begin{Lem}\label{Lem:special-w}
There exists an optimal solution of Problem~\ref{P1} with $i=\text{LB}$, denoted by  $\{\mathbf{w}_{\text{LB},l}^{\star}\}_{l=1}^{r_{\text{LB},1}}$, satisfying: 
\begin{align}
&\frac{\mathbf{w}_{\text{LB},l}^{\star}}{\norm{\mathbf{w}_{\text{LB},l}^{\star}}}=\frac{\mathbf{w}_{\text{LB},l'}^{\star}}{\norm{\mathbf{w}_{\text{LB},l'}^{\star}}},~l,l'\in\Delta\Upsilon_{g},g\in\mathcal{G},\label{eq:proof1-property2}\\
&\frac{\norm{\mathbf{w}_{\text{LB},l}^{\star}}^2}{\lVert\mathbf{w}_{\text{LB},r_{\text{LB},g}}^{\star}\rVert^2}\geq\frac{\Gamma_{\text{LB},l}2^{\sum_{j=l+1}^{r_{\text{LB},g}}{R_{\text{LB},j}}}}{\Gamma_{\text{LB},r_{\text{LB},g}}},~l\in\Delta\Upsilon_{g}\setminus\{r_{\text{LB},g}\},g\in\mathcal{G},\label{eq:proof1-norminequality}
\end{align}
where $\Delta\Upsilon_{g}=\{r_{\text{LB},g+1}+1,\ldots,r_{\text{LB},g}\}$.
\end{Lem}
\begin{Proof}
 See Appendix~A.
\end{Proof}

Recall that $\Delta\Upsilon_{g}$ represents the set of  all the layers in super-layer $g$. Thus, the property in \eqref{eq:proof1-property2} indicates that all layers in each super-layer have the same normalized optimal beamforming vector. The property in \eqref{eq:proof1-norminequality} indicates that in each super-layer, the ratio between the optimal power of any layer which is  not the highest layer and that of the highest layer has to be above a threshold.
By Lemma~\ref{Lem:special-w},  we have the following result.

\begin{Thm} The optimal values of Problem~\ref{P1} with $i=\text{LB}$ and $i=\text{QB}$ are the same, i.e., $P^{\star}_{\text{LB}}= P^{\star}_{\text{QB}}$.
\label{thm:P12}
\end{Thm}
\begin{Proof}
 See Appendix~B.
\end{Proof}

Recall that $\mathcal{S}_{\text{LB},\text{opt}}\subseteq\mathcal{S}_{\text{QB},\text{opt}}$. For the cases in $\mathcal{S}_{\text{LB,opt}}$, the obtained quality-based globally optimal solution (using Algorithm~\ref{alg:rank-reduction}) achieves the same total transmission power as the obtained layer-based globally optimal solution (using Algorithm~\ref{alg:rank-reduction}). For the cases in $\mathcal{S}_{\text{QB},\text{opt}}\setminus\mathcal{S}_{\text{LB},\text{opt}}$, we can still obtain a globally optimal solution of Problem~\ref{P1} for the quality-based scheme using Algorithm~\ref{alg:rank-reduction}, but may only obtain a locally optimal  solution of  Problem~\ref{P1} for the layer-based scheme using Algorithm~\ref{alg:penalty method}, i.e., the obtained quality-based optimal solution  achieves a lower total transmission power than  the obtained layer-based optimal solution for the cases in $\mathcal{S}_{\text{QB},\text{opt}}\setminus\mathcal{S}_{\text{LB},\text{opt}}$.


Except for the class of special cases in $\mathcal{S}_{\text{QB},\text{opt}}$, we cannot guarantee the minimum total transmission power of either formulation. To obtain more insights from the studied problems, we also compare  the optimal values of Problem~\ref{SDR1} with $i=\text{LB}$ and $i=\text{QB}$, i.e., lower bounds on the optimal values of  Problem~\ref{P1} with $i=\text{LB}$ and $i=\text{QB}$, which can be obtained using SDP.   By carefully exploring structures of the optimal solutions of Problem~\ref{SDR1} with $i=\text{LB}$ and $i=\text{QB}$, we have the following result.

\begin{Thm}The optimal values of Problem \ref{SDR1} with $i=\text{LB}$ and $i=\text{QB}$ are the same, i.e., $P_{\text{LB},\text{SDR}}^{\star}= P_{\text{QB},\text{SDR}}^{\star}$.
\label{thm:SDR12}
\end{Thm}
\begin{Proof}
 See Appendix~C.
\end{Proof}

Fig.~\ref{fig:simulation-compare-2md} (a) verifies Theorem~\ref{thm:P12} and Theorem~\ref{thm:SDR12} as well as the above discussion.

\begin{figure}[t]
\begin{center}
 \subfigure[\small{Total transmission power versus $N$. $r_{\text{LB},1}=5,r_{\text{LB},2}=3,r_{\text{LB},3}=1$.}]
 {\resizebox{5.4cm}{!}{\includegraphics{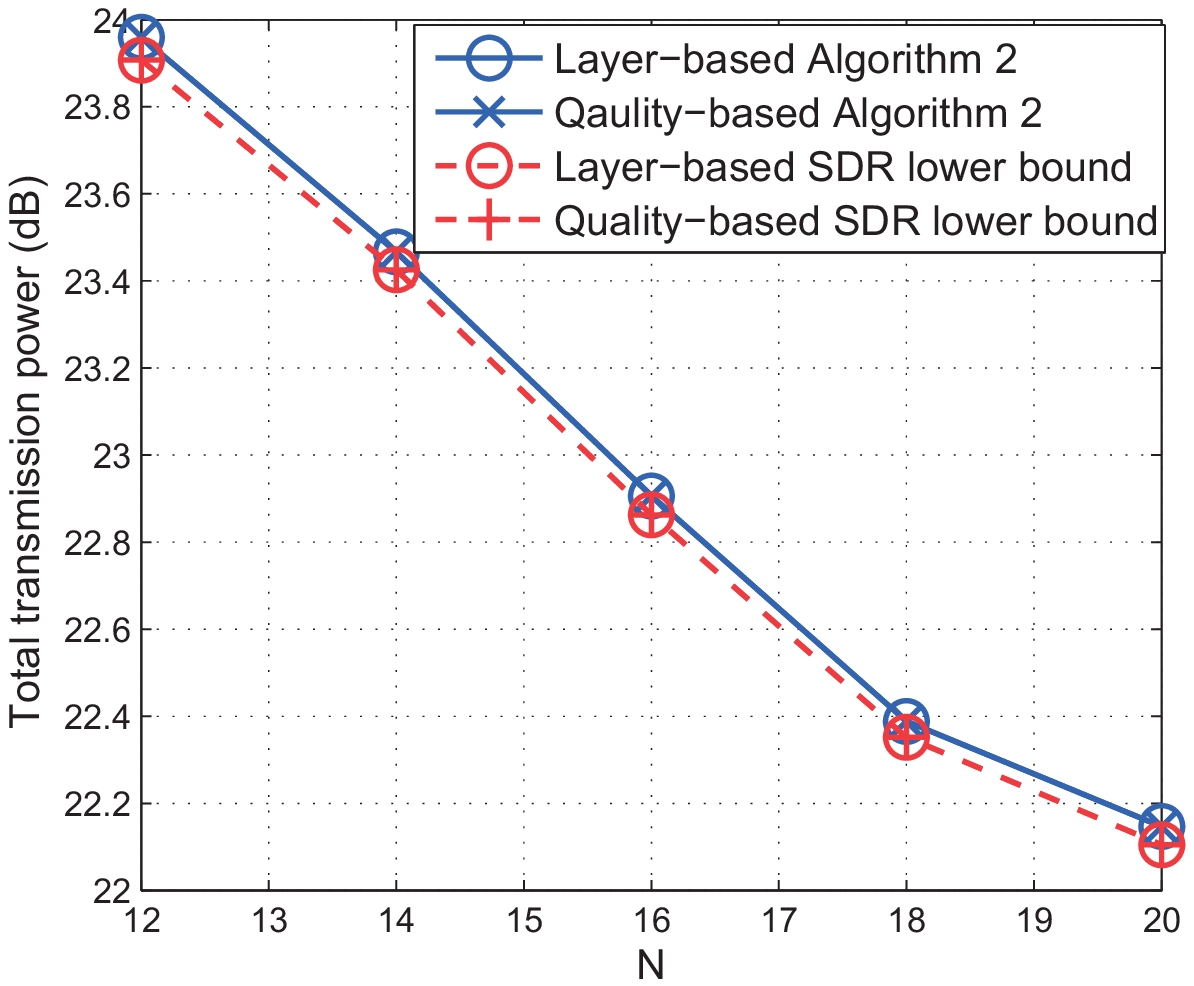}}}
 \subfigure[\small{Total running time versus $N$. $r_{\text{LB},1}=5,r_{\text{LB},2}=3,r_{\text{LB},3}=1$.}]
 {\resizebox{5.4cm}{!}{\includegraphics{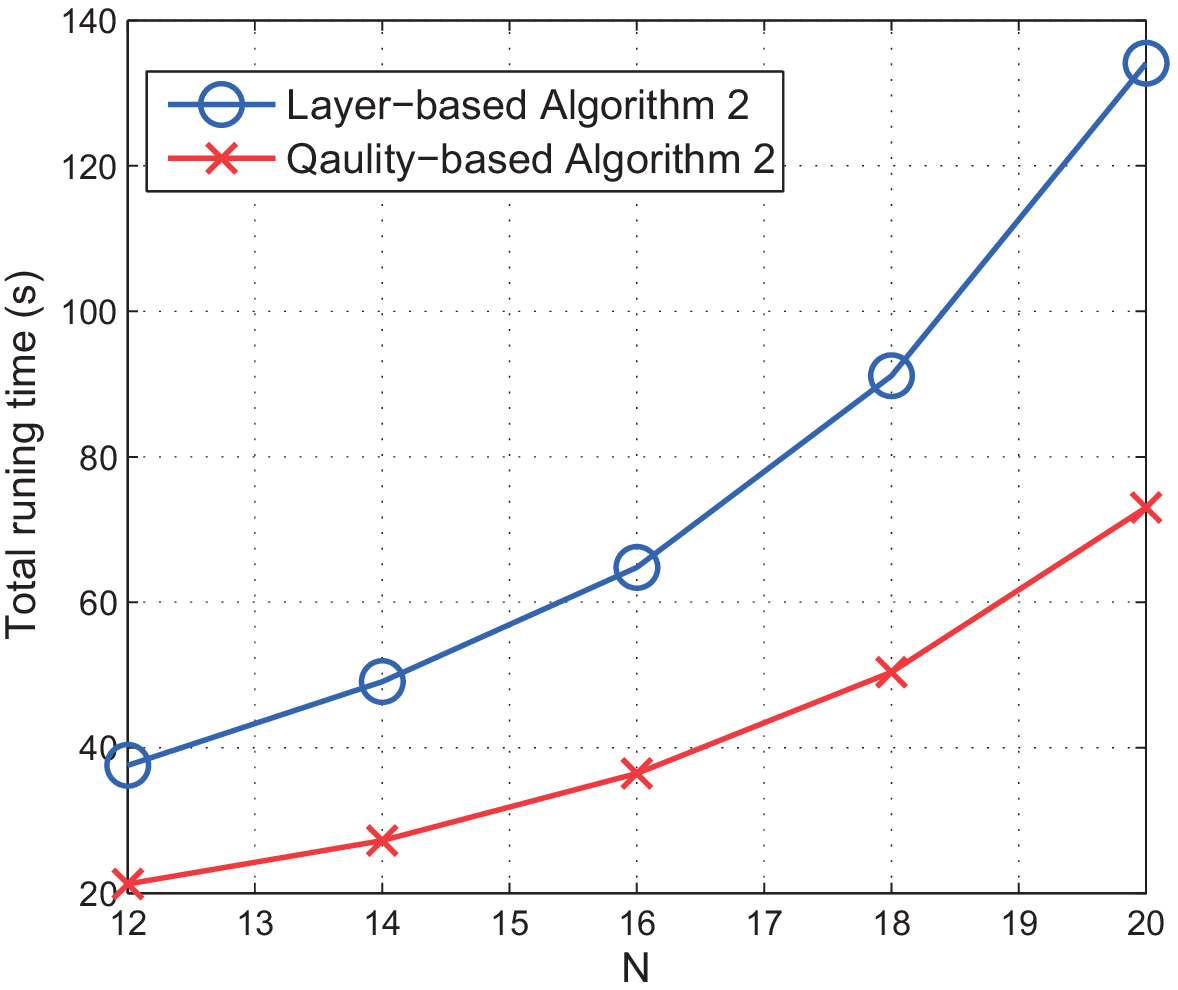}}}
 \subfigure[\small{Total running time versus $r_{\text{LB},1}$. $N=12,r_{\text{LB},2}=2,r_{\text{LB},3}=1$.}]
 {\resizebox{5.4cm}{!}{\includegraphics{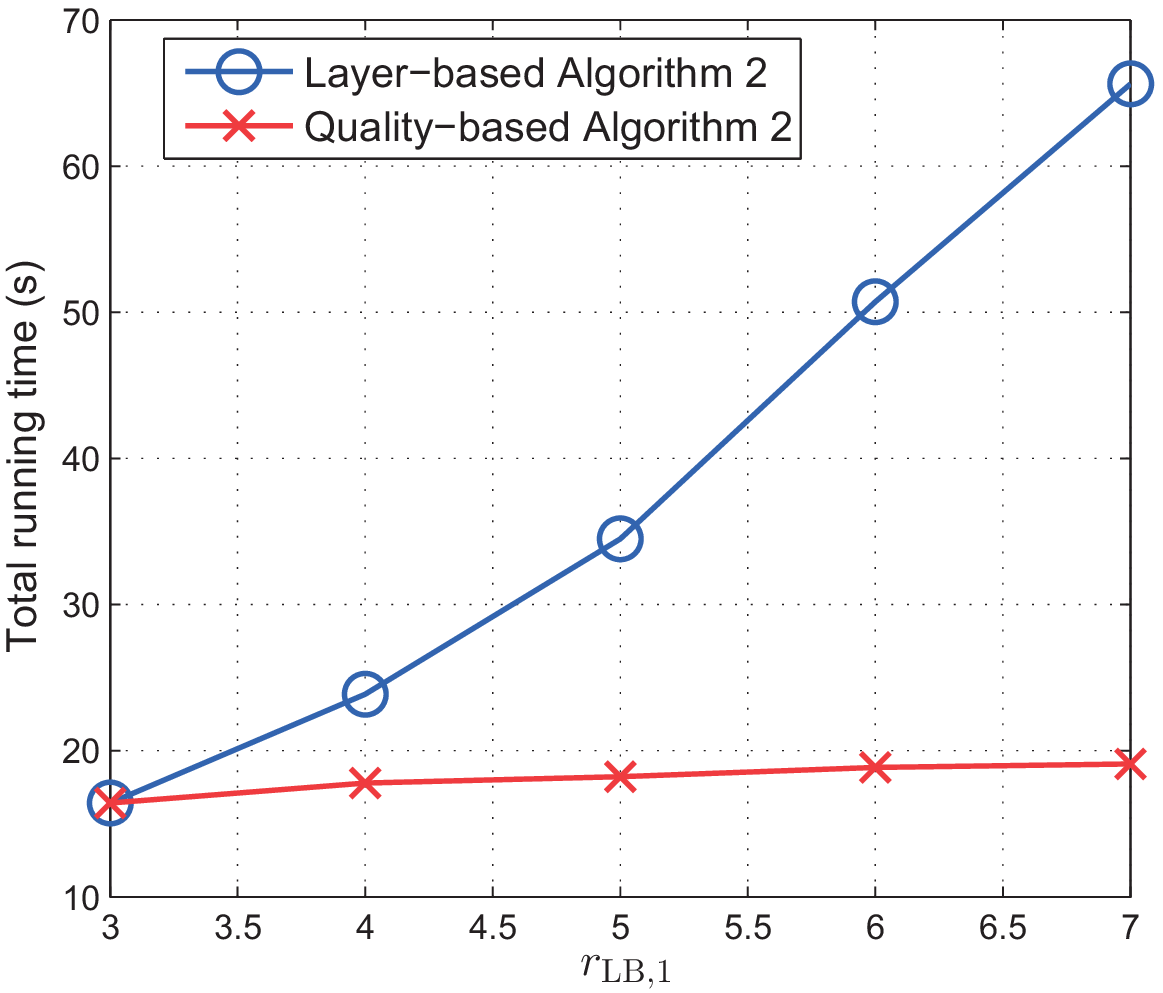}}}
 \end{center}
 \vspace{-0.5cm}
   \caption{\small{Comparison in total transmission power and total running time.~$G=3$, $U_1=3$, $U_2=3$, $U_3=3$, $R_{\text{LB},l}=2$, $\sigma_{u}^{2}=1$, $\mathbf{h}_{u}\sim\mathcal{CN}(0,\frac{1}{d_u^{\eta}}\mathbf{I}_{N})$, $d_u=2g-1$ and $\eta=2$ .}}
   \label{fig:simulation-compare-2md}
   \vspace{-1cm}
\end{figure}

\subsubsection{Computational Complexity}

In solving  Problem~\ref{P1} with $i=\text{LB}$ and $i=\text{QB}$, we consider their SDR problems, i.e., Problem~\ref{SDR1} with $i=\text{LB}$ and $i=\text{QB}$, respectively.
The computational complexities for solving Problem~\ref{SDR1} with $i=\text{LB}$ and $i=\text{QB}$ depend on the numbers of variables,  $L_{\text{LB},\text{max}}N^2$ and $GN^2$, respectively. Thus, in general, when $L_{\text{LB},\text{max}}>G$,  the computational complexity for solving Problem~\ref{P1} with $i=\text{LB}$ is higher than that for solving  Problem~\ref{P1} with $i=\text{QB}$. The computational complexity difference increases with $L_{\text{LB},\text{max}}-G$ and with $N$.  Fig.~\ref{fig:simulation-compare-2md}~(b) and Fig.~\ref{fig:simulation-compare-2md}~(c)  verify the above discussion.

\section{Total Utility Maximization}\label{section:layer selection}
In this section, we consider the total utility maximization in a longer time duration (more than 0.5~seconds) over which multiple groups of pictures (GOPs) are delivered, under a given maximum transmission power budget of the system. We first formulate the optimal joint layer selection and quality-based multi-quality multicast beamforming design problem to maximize the total utility of the system under a given maximum transmission power budget. Then, we propose  a greedy algorithm to obtain a near optimal solution.
\subsection{Problem Formulation}
Let $f(l)$ denote an arbitrary utility function of a user who decodes layers  $1,\ldots,l$, where $l\in\{1,\ldots,L\}$. Here, $f$  is assumed to be nonnegative and nondecreasing. 
Thus, $f(r_{\text{LB},g})$ represents the utility of a user in group $g\in\mathcal{G}$.  
Then, the total utility function of the system is given by $F(\mathbf{r}_{\text{LB}})\coloneqq\sum_{g=1}^{G}U_gf(r_{\text{LB},g})$, where  layer selection $\mathbf{r}_{\text{LB}}\in\mathcal{R}\coloneqq\{1,\ldots,L\}^G$.
Note that  videos are encoded in the unit of GOPs in practice. The transmission of one GOP (of duration $0.5-1$ second)  lasts more than $100$ time slots (each of $1-5$ milliseconds).
To guarantee the user experience, video quality should not change as frequently as channels and needs to remain constant for  a duration of one or multiple GOPs. In addition, recall that the optimal quality-based multi-quality multicast beamforming design achieves the same minimum total transmission power as the optimal layer-based multi-quality multicast beamforming design, with lower computational complexity.  Thus, we consider a reasonably long time duration, and   would like to study the optimal (constant) joint layer selection  and quality-based multi-quality multicast beamforming design to maximize the total utility under a given maximum transmission power budget, considering all  channel conditions.
To reduce the computational complexity, we divide the $U$ users in $\mathcal{U}$ into multiple groups according to their distances to the BS, and consider a single received video quality for all users in one group.
 With slight abuse of notation,
 let $G$ and $\mathcal{G}$ denote the total number of (distance-based) groups in the system and  the set of group indices, respectively, and let $\mathcal{U}_g~\subseteq\mathcal{U}$ denote the set of $U_g$  users in group $g\in\mathcal{G}$. Note that $\mathcal{U}_{g'}\cap\mathcal{U}_{g}=\emptyset$ for all $g',g\in\mathcal G$, $g'\neq g$ and  $\cup_{g\in\mathcal{G}}\mathcal{U}_{g}=\mathcal{U}$.


Let $\mathbf{H}\coloneqq(\mathbf{h}_u)_{u\in\mathcal{U}_{g},g\in\mathcal{G}}\in\mathcal{H}$ denote the global channel state of all users in the system, where $\mathcal{H}$ represents the space of global channel states.
To emphasize its dependence  on layer selection  $\mathbf{r}_{\text{LB}}$ and global channel state $\mathbf{H}$, we express the beamforming vector $\mathbf{w}_{\text{QB},l}$ as a function of $\mathbf{r}_{\text{LB}}$ and $\mathbf{H}$, i.e., $\mathbf{w}_{\text{QB},l}(\mathbf{r}_{\text{LB}},\mathbf{H})$.
Thus, similar to the SINR constraints in \eqref{constr:P1-sinr} of Problem~\ref{P1} for the quality-based scheme, we have the following  SINR constraints (for all $\mathbf{r}_{\text{LB}}\in\mathcal{R}$ and $\mathbf{H}\in\mathcal{H}$):
\begin{align}\label{constr:layer-sinr}
\frac{|\mathbf{h}_{u}^{H}(\mathbf{H})\mathbf{w}_{\text{QB},l}(\mathbf{r}_{\text{LB}},\mathbf{H})|^2}{\sum\limits_{k=l+1}^{L_{\text{QB},\text{max}}(\mathbf{r}_{\text{LB}})} {|\mathbf{h}_{u}^{H}(\mathbf{H})\mathbf{w}_{\text{QB},k}(\mathbf{r}_{\text{LB}},\mathbf{H})|^2+\sigma_{u}^{2}}}\geq \Gamma_{\text{QB},l}, \ u\in\mathcal{U}_{g}, l\in\Upsilon_{\text{QB},g}(\mathbf{r}_{\text{LB}}), g\in\mathcal{G}, \mathbf{r}_{\text{LB}}\in\mathcal{R}, \mathbf{H}\in\mathbf{\mathcal{H}},
\end{align}
where $L_{\text{QB},\text{max}}(\mathbf{r}_{\text{LB}})\coloneqq\max\{r_{\text{QB},1},\ldots,r_{\text{QB},G}\}$ denotes the total number of the layers needed to be transmitted from the BS to the users in the system, and $\Upsilon_{\text{QB},g}(\mathbf{r}_{\text{LB}})\coloneqq\{1,\ldots,r_{\text{QB},g}\}$ denotes the set of  layers received by the users in group $g\in\mathcal{G}$.
Let $\bar{P}$ denote the maximum transmission power budget  of the  system.
The total transmission power of the system should satisfy the following power constraint (for all $\mathbf{r}_{\text{LB}}\in\mathcal{R}$ and $\mathbf{H}\in\mathcal{H}$):
\begin{align}\label{constr:layer-power}
    &\sum_{l=1}^{L_{\text{QB},\text{max}}(\mathbf{r}_{\text{LB}})}\norm{\mathbf{w}_{\text{QB},l}(\mathbf{r}_{\text{LB}},\mathbf{H})}^2\leq \bar{P},~\mathbf{r}_{\text{LB}}\in\mathcal{R},\mathbf{H}\in\mathbf{\mathcal{H}}.
\end{align}
We now formulate the total utility maximization problem as follows.
\begin{Prob} [Total Utility Maximization]\label{Pnew}
\begin{align}
\max_{\mathbf{r}_{\text{LB}}\in\mathcal{R},\{\mathbf{w}_{\text{QB},l}(\mathbf{r}_{\text{LB}},\mathbf{H})\}_{l=1}^{L_{\text{QB},\text{max}}(\mathbf{r}_{\text{LB}})}} &\quad\sum_{g=1}^{G}U_gf(r_{\text{LB},g})\\
    \mathrm{s.t.} &\quad\eqref{constr:layer-sinr},~\eqref{constr:layer-power}.\nonumber
\end{align}
\end{Prob}
To solve Problem~\ref{Pnew}, for any given $\mathbf{r}_{\text{LB}}\in\mathcal{R}$, we need to find a feasible quality-based multicast beamforming design satisfying \eqref{constr:layer-sinr} and \eqref{constr:layer-power}, which is related to  Problem~\ref{P1} for the quality-based scheme and is NP-hard. Based on this connection, we transform Problem~\ref{Pnew} to an equivalent problem, which can be solved by utilizing the  solution of Problem~\ref{P1} for the quality-based scheme. First, we rewrite the optimal value of Problem~\ref{P1} with $i=\text{QB}$ for  given $\mathbf{r}_{\text{LB}}$ and $\mathbf{H}$ as $P_{\text{QB}}^{\star}(\mathbf{r}_{\text{LB}},\mathbf{H})$.
Note that $\mathbf{r}_{\text{LB}}\in\mathcal{R}$ is feasible if and only if the minimum total transmission power under the worst channel condition satisfies $\max\limits_{\mathbf{H}\in\mathbf{\mathcal{H}}}P_{\text{QB}}^{\star}(\mathbf{r}_{\text{LB}},\mathbf{H})\leq \bar{P}$.
Therefore,  based on Problem~\ref{P1} for the quality-based scheme, Problem~\ref{Pnew} can be equivalently transformed to the following problem.
\begin{Prob} [Equivalent Total Utility Maximization]\label{Pnew-equivalent}
\begin{align}
\max_{\mathbf{r}_{\text{LB}}\in\mathcal{R},\{\mathbf{w}_{\text{QB},l}(\mathbf{r}_{\text{LB}},\mathbf{H})\}_{l=1}^{L_{\text{QB},\text{max}}(\mathbf{r}_{\text{LB}})}} &\quad\sum_{g=1}^{G}U_gf(r_{\text{LB},g})\\
    \mathrm{s.t.}
    &\quad P_{\text{QB},\text{max}}^{\star}(\mathbf{r}_{\text{LB}})\leq \bar{P},~\mathbf{r}_{\text{LB}}\in\mathcal{R},
\end{align}
where $ P_{\text{QB},\text{max}}^{\star}(\mathbf{r}_{\text{LB}})\coloneqq\max\limits_{\mathbf{H}\in\mathbf{\mathcal{H}}}P_{\text{QB}}^{\star}(\mathbf{r}_{\text{LB}},\mathbf{H})$, and $P_{\text{QB}}^{\star}(\mathbf{r}_{\text{LB}},\mathbf{H})$ is given by Problem~\ref{P1} with $i=\text{QB}$.\nonumber
\end{Prob}

Problem~\ref{Pnew-equivalent} is still NP-hard. There are $L^G$ possible choices for $\mathbf{r}_{\text{LB}}\in\mathcal{R}$. For all $\mathbf{r}_{\text{LB}}\in\mathcal{R}$ and $\mathbf{H}\in\mathcal{H}$, we rewrite the SDR lower bound on $P_{\text{QB}}^{\star}(\mathbf{r}_{\text{LB}},\mathbf{H})$ and the locally optimal solution of Problem~\ref{P1} for the quality-based scheme as $P_{\text{QB},\text{SDR}}^{\star}(\mathbf{r}_{\text{LB}},\mathbf{H})$ and $P_{\text{QB}}^{\dagger}(\mathbf{r}_{\text{LB}},\mathbf{H})$,  respectively.
We can use  exhaustive search, as summarized in  Algorithm~\ref{alg:Opt Algorithm} for completeness, to solve Problem~\ref{Pnew-equivalent}.
Note that when $G=1$ and $U_1=U_2=1$, for any $\mathbf{r}_{\text{LB}}\in\mathcal{R}$, $(G,\mathbf{U},\mathbf{r}_{\text{LB}})$ satisfies \eqref{rank-reduction-constnum3}, i.e.,  $(G,\mathbf{U},\mathbf{r}_{\text{LB}})$ corresponds to a special case of Problem~\ref{P1} for the quality-based scheme, in which $P_{\text{QB},\text{SDR}}^{\star}(\mathbf{r}_{\text{LB}},\mathbf{H})=P_{\text{QB}}^{\star}(\mathbf{r}_{\text{LB}},\mathbf{H})$, as shown in Section~\ref{subsubsec:special-layer}. Therefore, if $G=1$ and $U_1=U_2=1$,  Algorithm~\ref{alg:Opt Algorithm} yields a globally optimal solution of Problem~\ref{Pnew-equivalent}; Otherwise,  Algorithm~\ref{alg:Opt Algorithm} provides a near optimal solution of Problem~\ref{Pnew-equivalent}.
\begin{algorithm}
    \caption{Exhaustive Search}
    \begin{multicols}{2}
     \begin{footnotesize}
    \begin{algorithmic}[1]
       \STATE  Sort $\mathbf{r}_{\text{LB}}\in\mathcal{R}$ by $F(\mathbf{r}_{\text{LB}})$ from highest to smallest, i.e., $F(\mathbf{r}_{\text{LB},(1)})\geq F(\mathbf{r}_{\text{LB},(2)})\geq\cdots\geq F(\mathbf{r}_{\text{LB},(L^G)})$;
           \STATE   Set $x=1$;
        \WHILE{$x\leq L^G$ and $\mathbf{r}_{\text{LB},(x)}\in\mathcal{R}$}
        \IF {$G==1, U_1==U_2==1$}
        \STATE For all $\mathbf{H}\in\mathcal{H}$, obtain $P_{\text{QB},\text{SDR}}^{\star}(\mathbf{r}_{\text{LB},(x)},\mathbf{H})$  by solving Problem~\ref{SDR1} with $i=\text{QB}$ using SDP, set $P_{\text{QB}}^{\star}(\mathbf{r}_{\text{LB},(x)},\mathbf{H})=P_{\text{QB},\text{SDR}}^{\star}(\mathbf{r}_{\text{LB},(x)},\mathbf{H})$, and compute $P_{\text{QB},\text{max}}(\mathbf{r}_{\text{LB},(x)})=\max\limits_{\mathbf{H}\in\mathbf{\mathcal{H}}}P_{\text{QB},\text{SDR}}^{\star}(\mathbf{r}_{\text{LB},(x)},\mathbf{H})$.
       \ELSE
       \STATE For all $\mathbf{H}\in\mathcal{H}$, obtain $P_{\text{QB}}^{\dagger}(\mathbf{r}_{\text{LB},(x)},\mathbf{H})$  by solving Problem~\ref{P1} with $i=\text{QB}$ using Algorithm~\ref{alg:penalty method}, and compute $P_{\text{QB},\text{max}}(\mathbf{r}_{\text{LB},(x)})=\max\limits_{\mathbf{H}\in\mathbf{\mathcal{H}}}P_{\text{QB},\text{SDR}}^{\dagger}(\mathbf{r}_{\text{LB},(x)},\mathbf{H})$
       \ENDIF
        \IF {$P_{\text{QB},\text{max}}(\mathbf{r}_{\text{LB},(x)})\leq\bar{P}$}
      \STATE Set $\mathbf{r}_{\text{ex}}=\mathbf{r}_{\text{LB},(x)}$;
      \STATE Break;
      \ELSE
      \STATE Set $x=x+1$;
      \ENDIF
      \ENDWHILE
      \STATE For all $\mathbf{H}\in\mathcal{H}$, obtain $\{\mathbf{w}_{\text{QB},l}^{\star}(\mathbf{r}_{\text{ex}},\mathbf{H})\}_{l=1}^{L_{\text{QB},\text{max}}(\mathbf{r}_{\text{LB}})}$ by solving Problem~\ref{SDR1} with $i=\text{QB}$ with $\mathbf{r}_{\text{ex}}$ using Algorithm~\ref{alg:rank-reduction}.
    \end{algorithmic}\label{alg:Opt Algorithm}
    \end{footnotesize}
    \end{multicols}
\end{algorithm}
\subsection{Greedy Algorithm}\label{subsec:ls-gready}
To reduce the computational complexity in solving Problem~\ref{Pnew-equivalent}, we introduce the following lemma.
\begin{Lem}\label{lemma2}
Let $ P_{\text{QB},\text{SDR},\text{max}}^{\star}(\mathbf{r}_{\text{LB}})\coloneqq\max\limits_{\mathbf{H}\in\mathbf{\mathcal{H}}}P_{\text{QB},\text{SDR}}^{\star}(\mathbf{r}_{\text{LB}},\mathbf{H})$.
If $\mathbf{r}_{\text{LB},1}\succcurlyeq\mathbf{r}_{\text{LB},2}$, then $P_{\text{QB},\text{SDR},\text{max}}^{\star}(\mathbf{r}_{\text{LB},1})\\
\geq P_{\text{QB},\text{SDR},\text{max}}^{\star}(\mathbf{r}_{\text{LB},2})$.
\end{Lem}
\begin{Proof}
 See Appendix~D.
\end{Proof}

Based on  Lemma~\ref{lemma2} and SDR, we develop a greedy algorithm to obtain a near optimal solution of Problem~\ref{Pnew-equivalent}, as summarized in Algorithm~\ref{Greedy Algorithm}.
For convenience, let $\Delta{P}_g(\mathbf{r}_{\text{LB}})\coloneqq P_{\text{QB},\text{SDR},\text{max}}^{\star}(\mathbf{r}_{\text{LB}})-P_{\text{QB},\text{SDR},\text{max}}^{\star}((r_{\text{LB},1},\ldots,r_{g-1},r_g-1,r_{g+1},\ldots,r_G))$ and $\Delta{F}_g(\mathbf{r}_{\text{LB}})\coloneqq F(\mathbf{r}_{\text{LB}})-F((r_{\text{LB},1},\ldots,r_{\text{LB},g-1},r_{\text{LB},g}-1,r_{\text{LB},g+1},\ldots,r_{\text{LB},G}))$ denote the power reduction and the utility reduction by changing $r_{\text{LB},g}$ to $r_{\text{LB},g}-1$ and keeping $r_{\text{LB},g'}$ unchanged for all $g'\neq g$ and $g,g'\in\mathcal{G}$, respectively.
Similarly, let $P_{\text{QB},\text{max}}^{\dagger}(\mathbf{r}_{\text{LB}})\coloneqq\max\limits_{\mathbf{H}\in\mathcal{H}}P_{\text{QB}}^{\dagger}(\mathbf{r}_{\text{LB}},\mathbf{H})$.
In Steps $3-11$, when $ P_{\text{QB},\text{SDR},\text{max}}^{\star}(\mathbf{r}_{\text{LB}})>\bar{P}$, i.e., $\mathbf{r}_{\text{LB}}$ is infeasible,
we remove the highest layer of group $g_{\text{min}}=\mathop{\arg\min}\limits_{g\in\{g\in\mathcal{G}|r_{\text{LB},g}\geq2\}}{\frac{\Delta{F}_g(\mathbf{r}_{\text{LB}})}{\Delta{P}_g(\mathbf{r}_{\text{LB}})}}$ iteratively.
Recall that the computational complexity for computing $ P_{\text{QB},\text{SDR},\text{max}}^{\star}(\mathbf{r}_{\text{LB}})$ is much lower than that for computing $P_{\text{QB},\text{max}}^{\dagger}(\mathbf{r}_{\text{LB}})$  and $ P_{\text{QB},\text{SDR},\text{max}}^{\star}(\mathbf{r}_{\text{LB}})\leq P_{\text{QB},\text{max}}^{\star}(\mathbf{r}_{\text{LB}})\leq P_{\text{QB},\text{max}}^{\dagger}(\mathbf{r}_{\text{LB}})$. Thus, to reduce the computational complexity for determining whether $\mathbf{r}_{\text{LB}}$ is infeasible, in Steps 4 and 5, we compute $ P_{\text{QB},\text{SDR},\text{max}}^{\star}(\mathbf{r}_{\text{LB}})$ and compare it with $\bar{P}$.
In Steps $12-20$, when $P_{\text{QB},\text{max}}^{\dagger}(\mathbf{r}_{\text{LB}})>\bar{P}$, similarly,
we remove the highest layer of group $g_{\text{min}}$ iteratively.
Note that $\mathbf{r}'_{\text{LB}}$ obtained by removing the highest layer of group $g_{\text{min}}$ from $\mathbf{r}_{\text{LB}}$ satisfies $\mathbf{r}_{\text{LB}}\succcurlyeq\mathbf{r}'_{\text{LB}}$.
By Lemma~\ref{lemma2}, we know that $ P_{\text{QB},\text{SDR},\text{max}}^{\star}(\mathbf{r}_{\text{LB}})\geq P_{\text{QB},\text{SDR},\text{max}}^{\star}(\mathbf{r}'_{\text{LB}})$. Thus, after Step 11, any $\mathbf{r}_{\text{LB}}$ satisfies $ P_{\text{QB},\text{SDR},\text{max}}^{\star}(\mathbf{r}_{\text{LB}})\leq\bar{P}$.
Therefore, in Steps 13 and 14, we directly compute $P_{\text{QB},\text{max}}^{\dagger}(\mathbf{r}_{\text{LB}})$ instead of $ P_{\text{QB},\text{SDR},\text{max}}^{\star}(\mathbf{r}_{\text{LB}})$, and compare it  with $\bar{P}$ to determine whether $\mathbf{r}_{\text{LB}}$ is infeasible.

\begin{algorithm}
    \caption{Greedy Algorithm}
    \begin{multicols}{2}
    \begin{footnotesize}
     \begin{algorithmic}[1]
           \STATE Set $r_{\text{LB},g}=L$, for all $g\in\mathcal{G}$;
           \STATE Set $flag=1$;
           \WHILE{$flag==1$}
           \STATE For all $\mathbf{H}\in\mathcal{H}$, obtain $P_{\text{QB},\text{SDR}}^{\star}(\mathbf{r}_{\text{LB}},\mathbf{H})$ by solving Problem~\ref{SDR1} with $i=\text{QB}$ using SDP, and compute $ P_{\text{QB},\text{SDR},\text{max}}^{\star}(\mathbf{r}_{\text{LB}})$;
           \IF{$ P_{\text{QB},\text{SDR},\text{max}}^{\star}(\mathbf{r}_{\text{LB}})>\bar{P}$}
           \STATE  Compute $\Delta{F}_g(\mathbf{r}_{\text{LB}})$ and $\Delta{P}_g(\mathbf{r}_{\text{LB}})$ for all $g\in\{g\in\mathcal{G}|r_g\geq2\}$;
            \STATE  Set $g_{\text{min}}=\mathop{\arg\min}\limits_{g\in\{g\in\mathcal{G}|r_{\text{LB},g}\geq2\}}{\frac{\Delta{F}_g(\mathbf{r}_{\text{LB}})}{\Delta{P}_g(\mathbf{r}_{\text{LB}})}}$
             and $r_{\text{LB},g_{\text{min}}}=r_{\text{LB},g_{\text{min}}}-1$;
            \ELSE
            \STATE Set $flag=0$;
           \ENDIF
           \ENDWHILE
           \WHILE{$flag==0$}
            \STATE For all $\mathbf{H}\in\mathcal{H}$, obtain $P_{\text{QB}}^{\dagger}(\mathbf{r}_{\text{LB}},\mathbf{H})$ by solving Problem~\ref{P1} with $i=\text{QB}$ using Algorithm~\ref{alg:penalty method}, and compute $P_{\text{QB},\text{max}}^{\dagger}(\mathbf{r}_{\text{LB}})$;
           \IF{$P_{\text{QB},\text{max}}^{\dagger}(\mathbf{r}_{\text{LB}})>\bar{P}$}
           \STATE  Compute $\Delta{F}_g(\mathbf{r}_{\text{LB}})$ and $\Delta{P}_g(\mathbf{r}_{\text{LB}})$ for all $g\in\{g\in\mathcal{G}|r_{\text{LB},g}\geq2\}$;
            \STATE  Set $g_{\text{min}}=\mathop{\arg\min}\limits_{g\in\{g\in\mathcal{G}|r_{\text{LB},g}\geq2\}}{\frac{\Delta{F}_g(\mathbf{r}_{\text{LB}})}{\Delta{P}_g(\mathbf{r}_{\text{LB}})}}$
             and $r_{\text{LB},g_{\text{min}}}=r_{\text{LB},g_{\text{min}}}-1$;
            \ELSE
            \STATE Break;
           \ENDIF
           \ENDWHILE
      \STATE Set $\mathbf{r}_{\text{greedy}}=\mathbf{r}_{\text{LB}}$;
       \STATE For all $\mathbf{H}\in\mathcal{H}$, obtain $\{\mathbf{w}_{\text{QB},l}^{\dagger}(\mathbf{r}_{\text{greedy}},\mathbf{H})\}_{l=1}^{L_{\text{QB},\text{max}}(\mathbf{r}_{\text{LB}})}$ by solving Problem~\ref{P1} with $i=\text{QB}$ and $\mathbf{r}_{\text{greedy}}$ using Algorithm~\ref{alg:penalty method}.
    \end{algorithmic}\label{Greedy Algorithm}
    \end{footnotesize}
    \end{multicols}
\end{algorithm}

\begin{figure}[t]
\begin{center}
 \subfigure[\small{Total utility versus $\bar{P}$.}]
 {\resizebox{5.6cm}{!}{\includegraphics{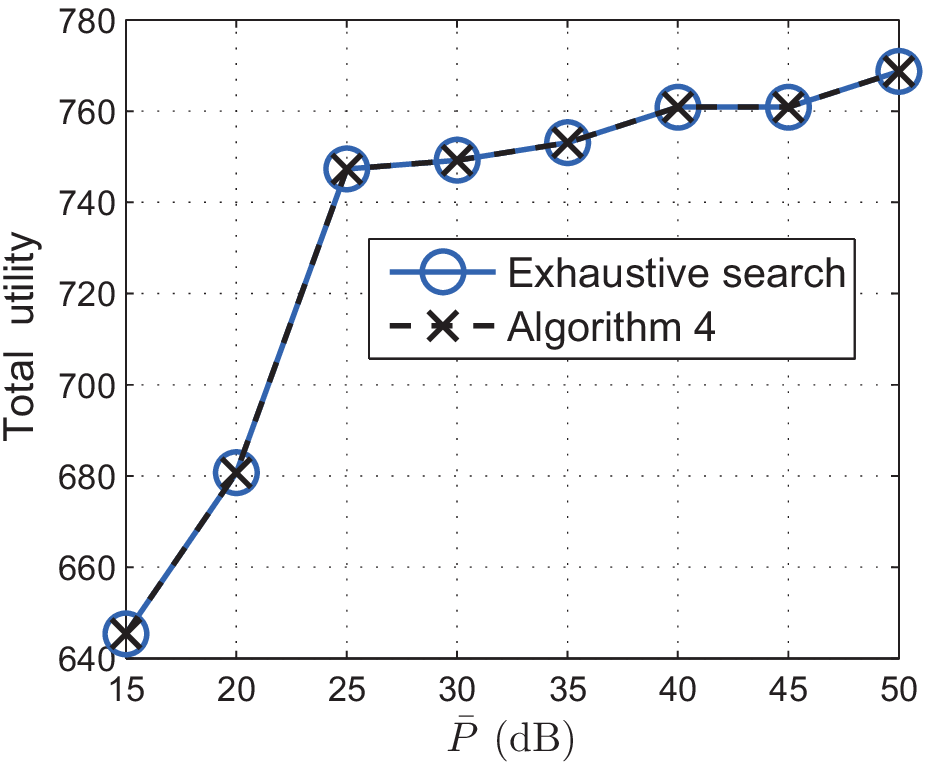}}}
 \subfigure[\small{Total running time versus $\bar{P}$. }]
 {\resizebox{5.4cm}{!}{\includegraphics{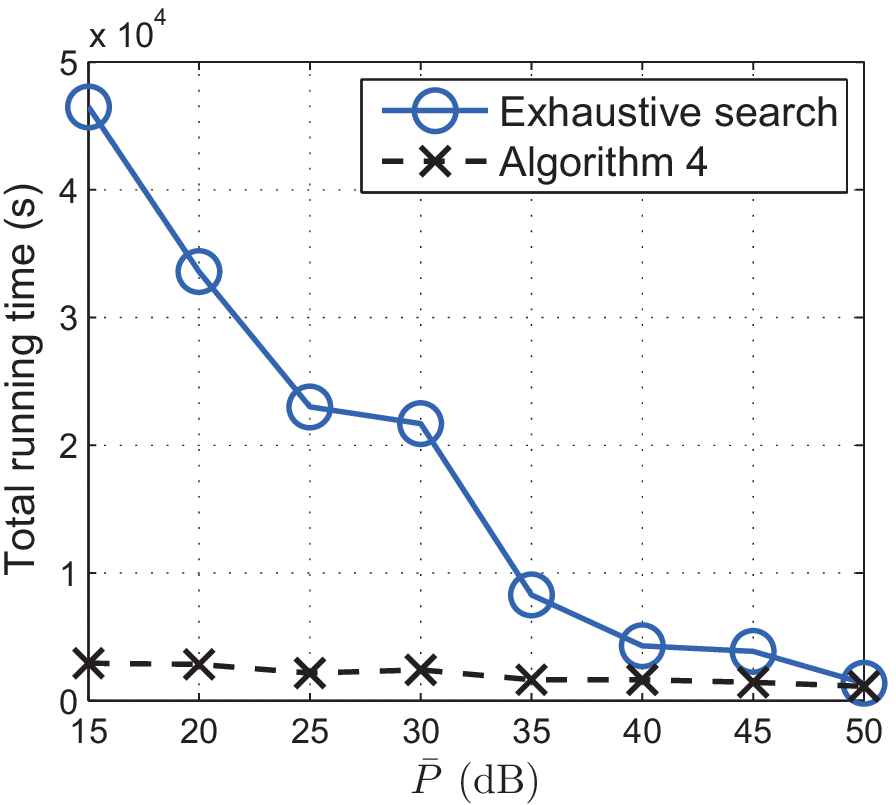}}}
 \end{center}
 \vspace*{-0.5cm}
   \caption{\small{Comparison in total utility and total running time. $G=4$, $U_g=5$, $\sigma_{u}^{2}=1$, $\mathbf{h}_u\sim\mathcal{CN}(0,\frac{1}{d_u^{\eta}}\mathbf{I})$, $d_u=2g-1$ and $\eta=2$. The parameters related to the video can be found in Table~\ref{table:parameter}.}}
   \label{fig:algorithm34}
\end{figure}

Now, for the case of $G=4$ and $U_g=5$ for all $g\in\mathcal{G}$, we compare the performance of Algorithm~\ref{alg:Opt Algorithm} and Algorithm~\ref{Greedy Algorithm} using numerical results.
From Fig.~\ref{fig:algorithm34} (a), we observe that the total utility achieved by Algorithm~\ref{Greedy Algorithm}  is quite close to  that achieved by Algorithm~\ref{alg:Opt Algorithm}.
From Fig.~\ref{fig:algorithm34} (b), we observe that the total running time of Algorithm~\ref{alg:Opt Algorithm}  decreases with the maximum  transmission power budget, and is always longer than that of  Algorithm~\ref{Greedy Algorithm}.
\begin{table}[t]
\caption{Parameter setting of the video \cite{layers16}.}\label{table:parameter}
\vspace{-0.9cm}
\scriptsize
\begin{center}
\begin{tabular}{|l|l|l|}
\hline
\makecell[c]{Layer $l$} &\makecell[c]{Utility $f(l)$ ($\text{PSNR}_l$)} &\makecell[c]{Data rate $R_{\text{LB},l}$ (bps/Hz)} \\
\hline
1  &28.1496 &0.7447\\
\hline
2  &30.6066 &0.9358\\
\hline
3  &37.2694 &2.9765 \\
\hline
4 &37.6534 &3.3320\\
\hline
5 &39.2136  &7.1040\\
\hline
\end{tabular}
\vspace{-1.3cm}
\end{center}
\end{table}

\section{Simulation Results}
In this section, we compare the proposed solutions in Sections~\ref{subsec:power-optimal} and \ref{subsec:ls-gready} with some baselines using numerical results.
In the simulation, we use JSVM as the SVC encoder and MEPG video sequence $Kendo$ as the video source \cite{layers16}.
The video is encoded into five layers with the resulted Peak Signal-to-Noise Ratio (PSNR) for each quality level and data rate for each layer given by Table~\ref{table:parameter}.    Let $\text{PSNR}_l$ denote the PSNR value for quality level $l$, which is determined by the SVC-encoded video and the ground truth (original video) \cite{layers16}. We choose $f(l)=\text{PSNR}_l$.
\subsection{Multi-quality Multicast Beamforming Design}\label{section:simulation-A}
\begin{figure}[t]
\begin{center}
 \subfigure[\small{Total transmission power versus quality of group 2. $G=2$.}]
 {\resizebox{5.4cm}{!}{\includegraphics{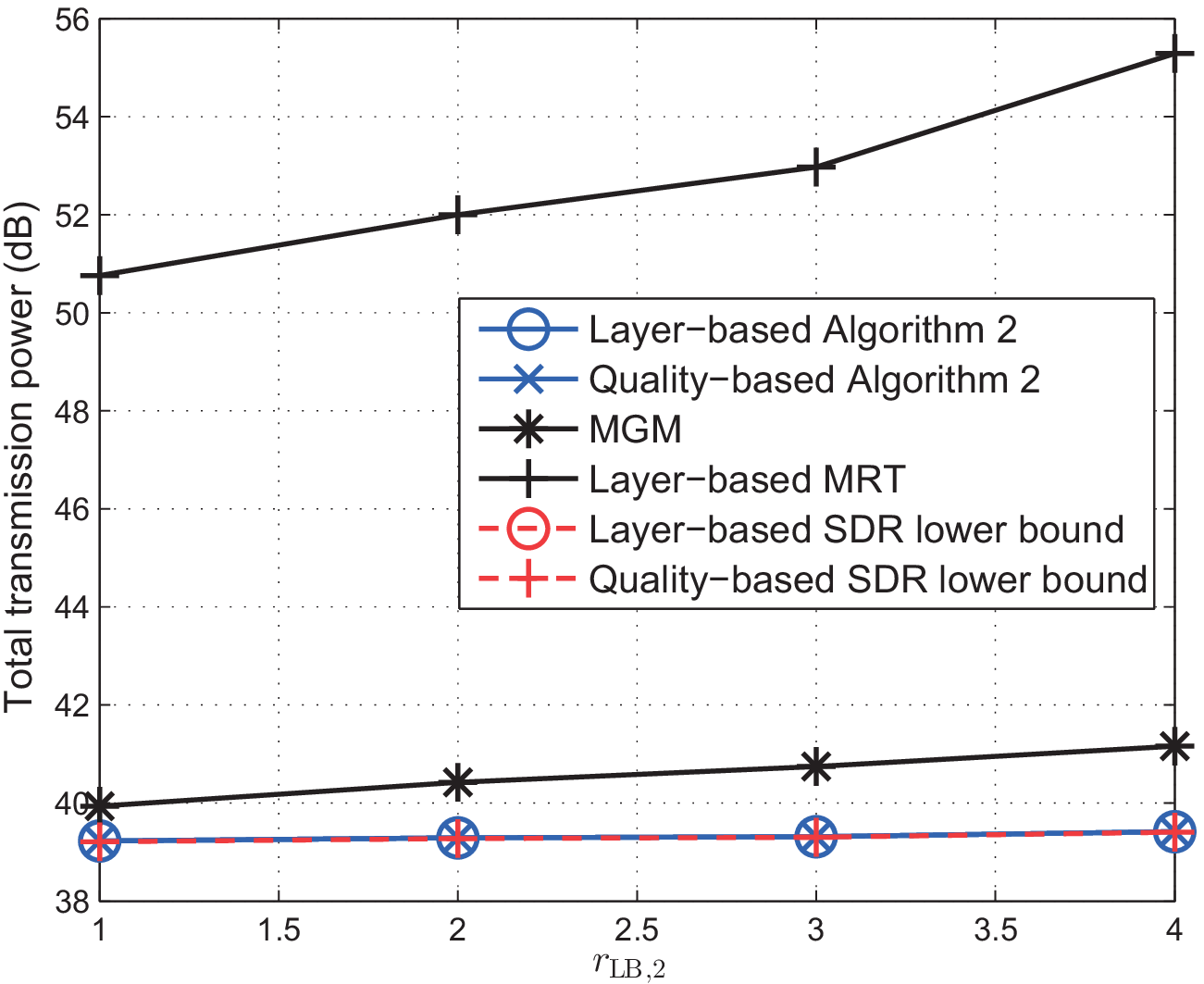}}}
 \subfigure[\small{Total transmission power versus total number of the groups. $r_{\text{LB},g}=6-g$.}]
 {\resizebox{5.4cm}{!}{\includegraphics{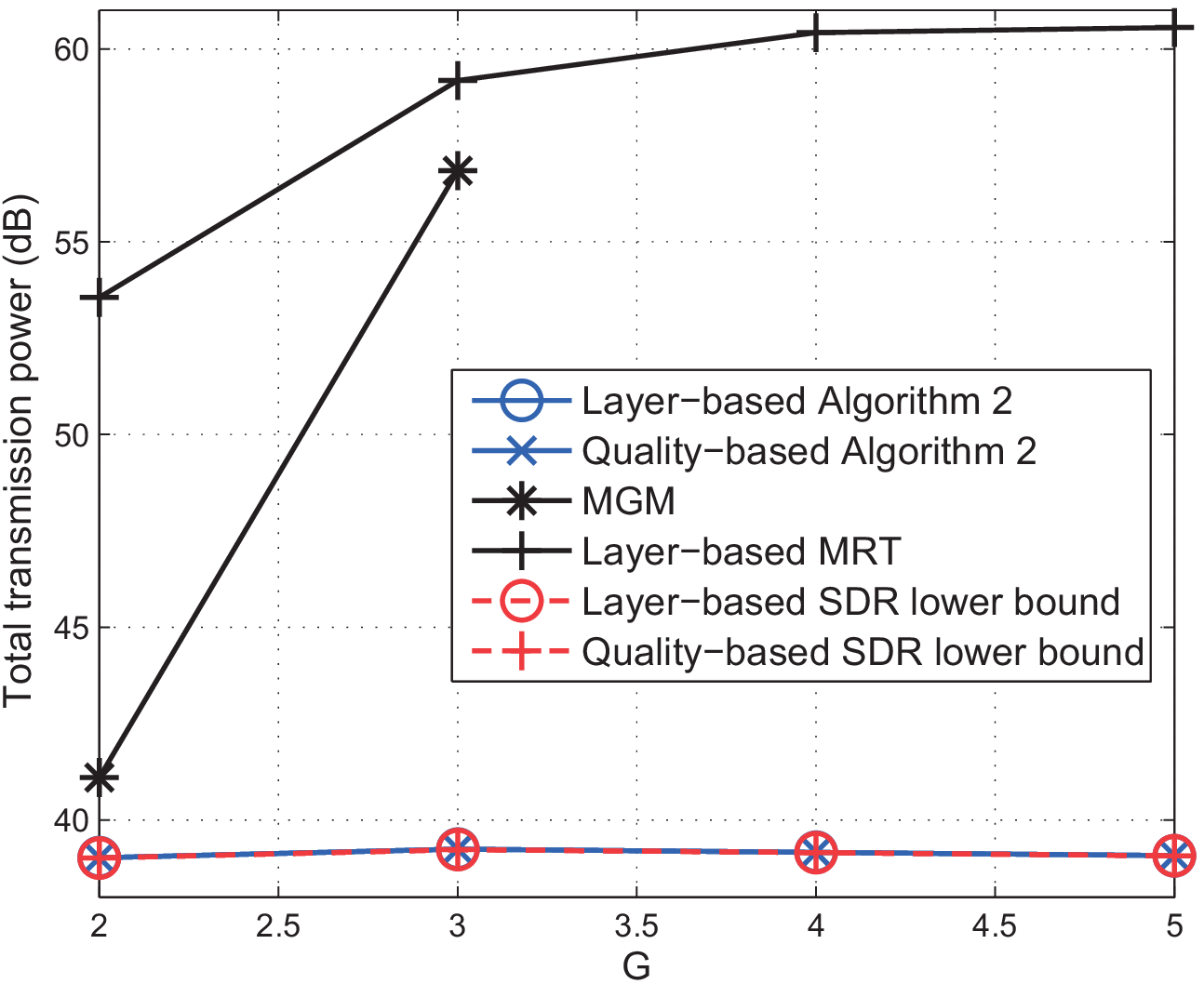}}}
  \subfigure[\small{Total transmission power versus number of antennas  at the BS. $G=3$, $r_{\text{LB},2}=3$, $r_{\text{LB},3}=1$.}]
 {\resizebox{5.4cm}{!}{\includegraphics{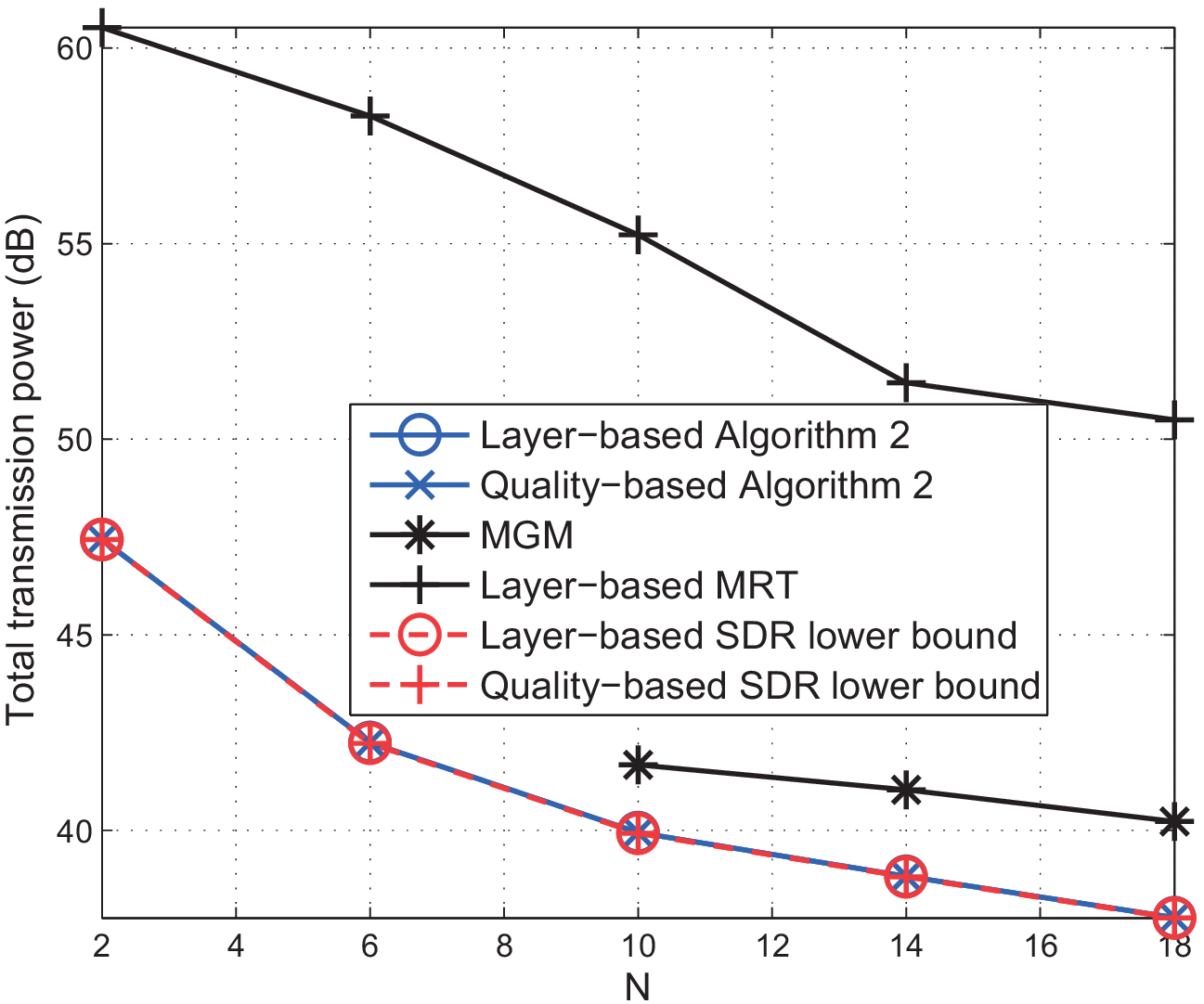}}}
 \end{center}
 \vspace*{-0.5cm}
   \caption{\small{Total transmission power versus different system parameters.~$N=12$, $r_{\text{LB},1}=5$, $U_g=4$, $\sigma_{u}^{2}=1$, $\mathbf{h}_u\sim\mathcal{CN}(0,\frac{1}{d_u^{\eta}}\mathbf{I})$, $d_u=2g-1$ and $\eta=2$.
   In (b) and (c), when $G>3$ or $N<10$, the multi-group multicast baseline is always infeasible, indicating that it has a strictly smaller feasibility region than the two proposed optimal solutions.}}
   \label{fig:simulation-result1}
   \vspace*{-1cm}
\end{figure}
In this section, we compare the proposed layer-based and quality-based optimal solutions of Problem~\ref{P1} with $i=\text{LB}$ and $i=\text{QB}$ (both obtained using Algorithm~\ref{alg:penalty method}) with their corresponding SDR lower bounds and  two baselines. The first baseline treats the video streams of different qualities for different groups as  independent  messages and applies the existing solution   for multi-group multicast in \cite{luo2008}. This baseline  is referred to as multi-group multicast (MGM).  The second baseline  treats the users requiring the same layer as a multi-antenna super-user, adopts the normalized beamformer for each layer  according to  maximum ratio transmission (MRT) for the channel power gain matrix of the super-user, and   minimizes the total transmission power by optimizing the power of each beamformer. This baseline   is referred to as  layer-based MRT.

Fig.~\ref{fig:simulation-result1} illustrates the total transmission power versus different system parameters. From Fig.~\ref{fig:simulation-result1}, we can observe that the two proposed optimal solutions of Problem~\ref{P1} perform similarly and achieve close-to-optimal total transmission power (as the achieved total transmission powers are close to the corresponding SDR lower bounds). Besides, the two proposed optimal solutions  significantly outperform the two baselines.
Their performance gains  over the MGM baseline  come from the fact that the proposed multi-quality multicast beamforming schemes utilize SVC. Their performance gains  over the layer-based MRT baseline are due to the fact that the proposed schemes provide a higher flexibility in steering the beamformer to reduce the power consumption under the quality requirements.

\subsection{Total Utility Maximization}
In this part, we compare the proposed joint layer selection and quality-based multi-quality multicast beamforming design with two baselines, which correspond to the greedy solutions of the total utility maximization problems, under the MGM baseline and the layer-based MRT baseline considered in Section~\ref{section:simulation-A}, respectively. The two baselines are referred to as MGM-based layer selection and MRT-based layer selection, respectively.
\begin{figure}[t]
\begin{center}
 \subfigure[\small{Total utility versus  maximum transmission power budget. $N=12$,  $U_g=5$.}]
 {\resizebox{5.4cm}{!}{\includegraphics{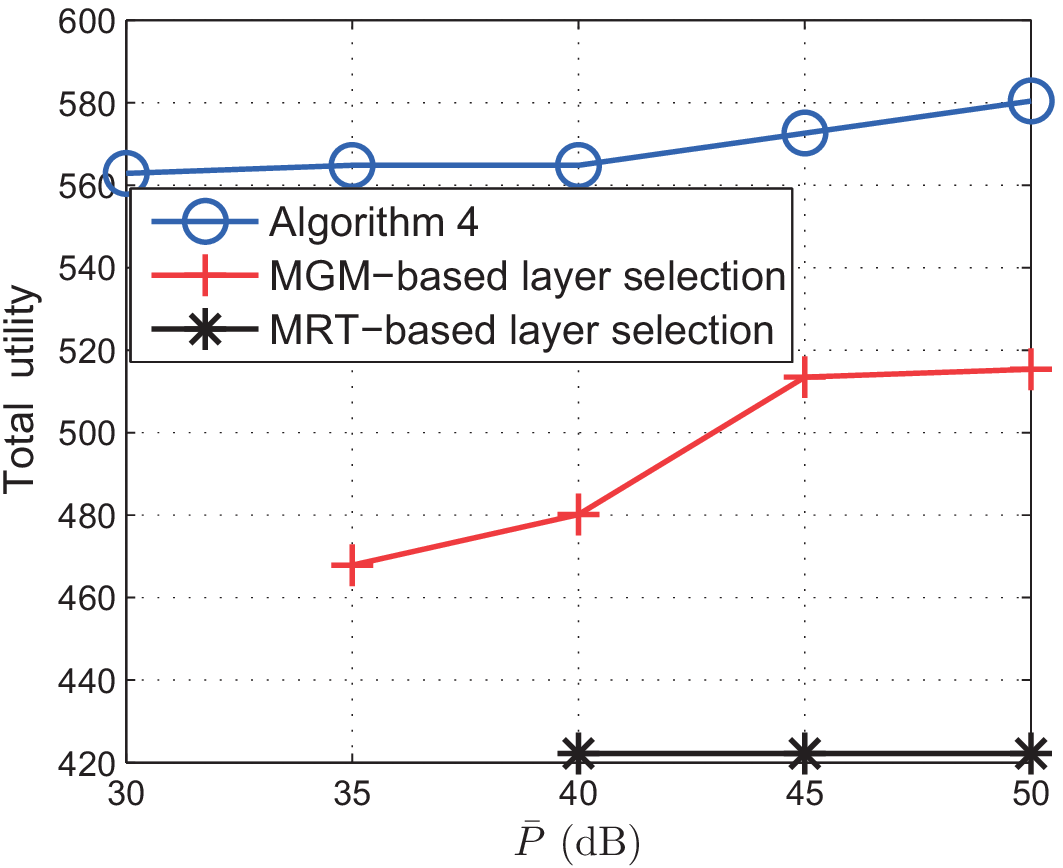}}}
 \subfigure[\small{Total utility versus  number of antennas at the BS. $\bar{P}=35$ dB,  $U_g=5$.}]
 {\resizebox{5.4cm}{!}{\includegraphics{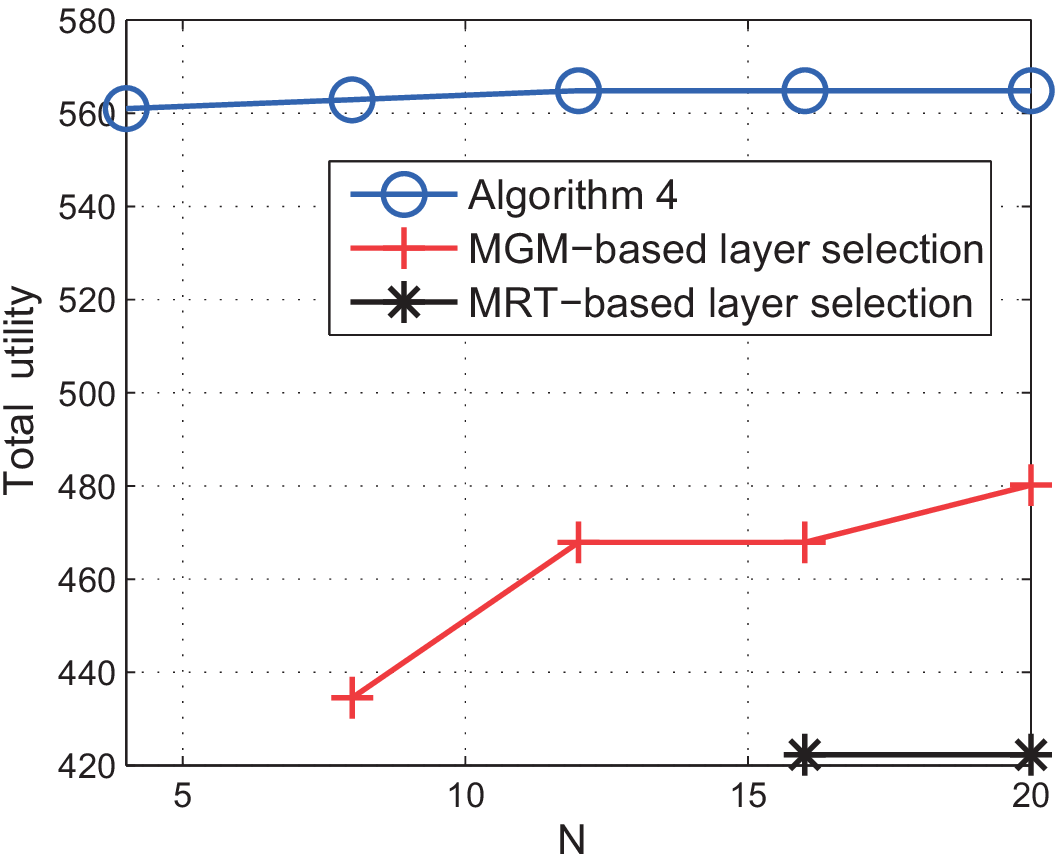}}}
  \subfigure[\small{Total utility versus  number of users in each group. $\bar{P}=40$ dB, $N=12$.}]
 {\resizebox{5.4cm}{!}{\includegraphics{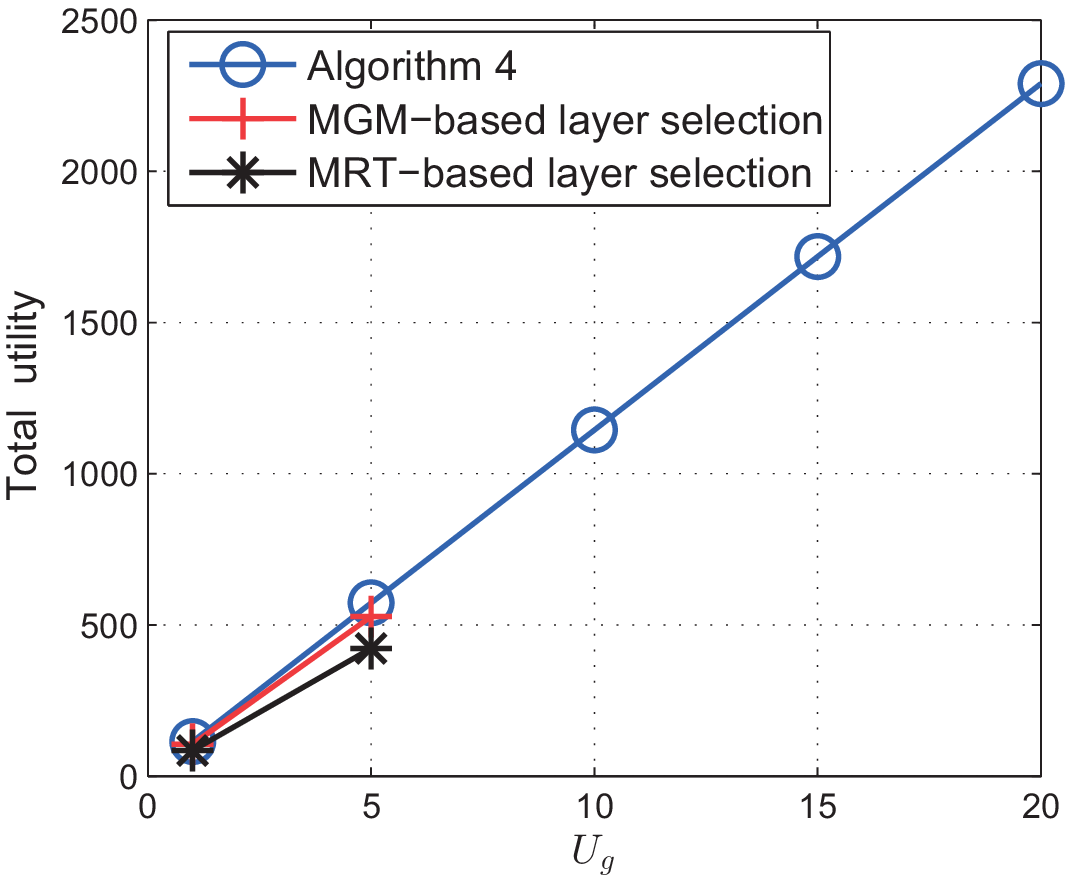}}}
 \end{center}
 \vspace*{-0.6cm}
   \caption{\small{Total utility versus  different system parameters. $G=3$, $\sigma_{u}^{2}=1$. $U_g$ are the same, where $g\in\mathcal{G}$. $\mathcal{H}$ contains 1000 global channel states randomly chosen according to $\mathbf{h}_u\sim\mathcal{CN}(0,\frac{1}{d_u^{\eta}}\mathbf{I})$, $d_u=2g-1$ and $\eta=2$. When $\bar{P}$ and $N$ are small and $\bar{U}$ is large, the two baselines are always infeasible.}}
   \label{fig:simulation-result2}
   \vspace*{-1.1cm}
\end{figure}
Fig.~\ref{fig:simulation-result2} illustrates the total utility versus different system parameters. From Fig.~\ref{fig:simulation-result2}, we can observe that the proposed solution obtained by Algorithm~\ref{Greedy Algorithm} significantly outperforms the two baselines. This is due to the fact that the proposed quality-based multi-quality multicast beamforming scheme utilizes SVC, and fully exploits the spatial degrees of freedom in multicasting offered by multiple antennas.
 It is worth noting that in Fig~\ref{fig:simulation-result2} (c), the total utility achieved by Algorithm~\ref{Greedy Algorithm} is proportional to the number of users in each group (i.e., the greedy solution does not change with the number of users in each group), as  the total transmission power achieved by the proposed quality-based optimal solution increases slowly with the number of users in each group.
\section{Conclusion}

In this paper, we first proposed  two beamforming schemes for layer-based multi-quality multicast and quality-based multi-quality multicast.
For each scheme, we formulated  the minimum power multi-quality multicast beamforming problem under given quality requirements of all users, and obtained a globally optimal solution in a class of special cases as well as a locally optimal solution for the general case.
Then, we showed that the proposed quality-based  optimal solution achieves lower total transmission power than  the proposed layer-based optimal solution in some  cases and the same total transmission power in other cases. However, the quality-based optimal solution incurs a much lower computational complexity than the  layer-based optimal solution. Next, we considered the optimal joint layer selection and quality-based multi-quality multicast beamforming design to maximize the total utility under a given maximum transmission power budget. We developed  a greedy algorithm to obtain a near optimal solution.
Finally, numerical results  unveiled the power
savings and total utility gains enabled by the proposed solutions over existing solutions.

\section{Acknowledgements}
We are grateful to Jenq-Neng Hwang for useful discussions and comments.
\section*{Appendix A: Proof of Lemma 1}\label{app:lemma 1}
 By \eqref{constr:P1-sinr}, we have
\begin{align}\label{lem:proof1-P1-sinr}
\norm{\mathbf{w}_{\text{LB},l}}^2\geq\frac{\Gamma_{\text{LB},l}\left(\sum\limits_{k=l+1}^{r_{\text{LB},1}} {|\mathbf{h}_{u}^{H}\mathbf{w}_{\text{LB},k}|^2+\sigma_{u}^{2}}\right)}{|\mathbf{h}_{u}^H\frac{\mathbf{w}_{\text{LB},l}}{\norm{\mathbf{w}_{\text{LB},l}}}|^2},~l\in\Upsilon_{\text{LB},g},u\in\mathcal{U}_g,g\in\mathcal{G}.
\end{align}
Thus, for any optimal solution $\{\mathbf{w}_{\text{LB},l}^{\star}\}_{l=1}^{r_{\text{LB},1}}$, we have
\begin{align}\label{lem:proof1-power}
\norm{\mathbf{w}_{\text{LB},l}^{\star}}^2=\Gamma_{\text{LB},l}\Psi_{l}\left(\{\mathbf{w}_{\text{LB},k}^{\star}\}_{k=l+1}^{r_{\text{LB},1}}\right),~l\in\Upsilon_{\text{LB},\text{max}},
\end{align}
where
\begin{align}
(\text{P}_l)\left\{
\begin{aligned}
\Psi_{l}\left(\{\mathbf{w}_{\text{LB},k}^{\star}\}_{k=l+1}^{r_{\text{LB},1}}\right)\coloneqq\min_{\mathbf{\bar{w}}\in\mathbb{C}^{N\times1}}&\max_{u\in\bigcup_{g\in\{g|l\in\Upsilon_{\text{LB},g}\}}\mathcal{U}_g}\left\{\frac{\sum\limits_{k=l+1}^{r_{\text{LB},1}} {|\mathbf{h}_{u}^{H}\mathbf{w}_{\text{LB},k}^{\star}|^2+\sigma_{u}^{2}}}{|\mathbf{h}_{u}^H\mathbf{\bar{w}}|^2}\right\}\\
\mathrm{s.t.}\quad & \norm{\mathbf{\bar{w}}}=1,~|\mathbf{h}_{u}^H\mathbf{\bar{w}}|=0,u\in\bigcup_{g\in\{g|l\notin\Upsilon_{\text{LB},g}\}}\mathcal{U}_g.
\end{aligned}
\right.
\end{align}
Thus, the normalized vector of $\mathbf{w}_{\text{LB},l}^{\star}$ denoted by $\mathbf{\bar{w}}_{\text{LB},l}^{\star}\coloneqq \frac{\mathbf{w}_{\text{LB},l}^{\star}}{\norm{\mathbf{w}_{\text{LB},l}^{\star}}}$ is an optimal solution of $(\text{P}_l)$, where $l\in\Upsilon_{\text{LB},\text{max}}$.

Now, according to the increasing order of $g$, we obtain $\{\mathbf{w}_{\text{LB},l}^{\star}\}_{l=r_{\text{LB},g+1}+1}^{r_{\text{LB},g}}$ in $r_{\text{LB},g}-r_{\text{LB},g+1}$ steps for each $g\in\mathcal{G}$. In the first step,   we obtain $\mathbf{w}_{\text{LB},r_{\text{LB},g}}^{\star}=\norm{\mathbf{w}_{\text{LB},r_{\text{LB},g}}^{\star}}\mathbf{\bar{w}}_{\text{LB},r_{\text{LB},g}}^{\star}$. Specifically, we obtain $\norm{\mathbf{w}_{\text{LB},r_{\text{LB},g}}^{\star}}$ by \eqref{lem:proof1-power} and obtain $\mathbf{\bar{w}}_{\text{LB},r_{\text{LB},g}}^{\star}$ by solving $(\text{P}_{r_{\text{LB},g}})$. In the $k$th step, where $k\in\{2,\ldots,r_{\text{LB},g}-r_{\text{LB},g+1}\}$, we obtain $\mathbf{w}_{\text{LB},l}^{\star}=\norm{\mathbf{w}_{\text{LB},l}^{\star}}\mathbf{\bar{w}}_{\text{LB},l}^{\star}$, where $l=r_{\text{LB},g}-k+1$. Specifically, we obtain $\norm{\mathbf{w}_{\text{LB},l}^{\star}}$ by \eqref{lem:proof1-power} and choose $\mathbf{\bar{w}}_{\text{LB},l}^{\star}=\mathbf{\bar{w}}_{\text{LB},l+1}^{\star}$.
First, we show that $\mathbf{\bar{w}}_{\text{LB},l+1}^{\star}$ is a feasible solution of $(\text{P}_{l})$. Since $\mathbf{\bar{w}}_{\text{LB},l+1}^{\star}$ is an optimal solution of  $(\text{P}_{l+1})$, it satisfies
\begin{align}
|\mathbf{h}_{u}^H\mathbf{\bar{w}}_{\text{LB},l+1}^{\star}|=0,~u\in\bigcup_{g\in\{g|l+1\notin\Upsilon_{\text{LB},g}\}}\mathcal{U}_g.
\end{align}\label{lem:constr-addtional-1}
By noting that $l\in\Upsilon_{\text{LB},g}-\Upsilon_{\text{LB},g+1}$ and $l+1\in\Upsilon_{\text{LB},g}-\Upsilon_{\text{LB},g+1}$, we have $\{g|l+1\notin\Upsilon_{\text{LB},g}\}=\{g|l\notin\Upsilon_{\text{LB},g}\}$. Thus, $\mathbf{\bar{w}}_{\text{LB},l+1}^{\star}$ also satisfies
\begin{align}
|\mathbf{h}_{u}^H\mathbf{\bar{w}}_{\text{LB},l+1}^{\star}|=0,~u\in\bigcup_{g\in\{g|l\notin\Upsilon_{\text{LB},g}\}}\mathcal{U}_g.
\end{align}\label{lem:constr-addtional-2}
Therefore, the normalized vector $\mathbf{\bar{w}}_{\text{LB},l+1}^{\star}$  is a feasible solution of $(\text{P}_{l})$.
Next, we can show that $\mathbf{\bar{w}}_{\text{LB},l+1}^{\star}$ is also an optimal solution of $(\text{P}_l)$, by contradiction \cite{fulllamma}.
Finally, we show that the obtained optimal solution $\{\mathbf{w}_{\text{LB},l}^{\star}\}_{l=1}^{r_{\text{LB},1}}$ of Problem~\ref{P1}  satisfies the two properties in \eqref{eq:proof1-property2} and  \eqref{eq:proof1-norminequality}. By the construction of $\{\mathbf{w}_{\text{LB},l}^{\star}\}_{l=1}^{r_{\text{LB},1}}$, we can see that \eqref{eq:proof1-property2}  holds.
In addition,
by \eqref{constr:P1-sinr},   we have
\begin{align}\label{eq:proof1-inequality1}
|\mathbf{h}_{u}^H\mathbf{w}_{\text{LB},l}^{\star}|^2\geq\Gamma_{\text{LB},l}\left(\sum\limits_{k=l+1}^{r_{\text{LB},1}} {|\mathbf{h}_{u}^{H}\mathbf{w}_{\text{LB},k}^{\star}|^2+\sigma_{u}^{2}}\right),~u\in\bigcup_{g\in\{g|l\in\Upsilon_{\text{LB},g}\}}\mathcal{U}_g,l\in\Upsilon_{\text{LB},\text{max}}.
\end{align}
From \eqref{lem:proof1-power} and \eqref{eq:proof1-inequality1}, for all $l\in\{r_{\text{LB},g+1}+1,\ldots,r_{\text{LB},g}-1\},g\in\mathcal{G}$,  we have
\begin{align}\label{eq:proof1-inequality2}
\norm{\mathbf{w}_{\text{LB},l}^{\star}}^2\geq&\Gamma_{\text{LB},l} 2^{\sum_{j=l+1}^{r_{\text{LB},g}}{R_{\text{LB},j}}}\Psi_{r_{\text{LB},g}}\left(\{\mathbf{w}_{\text{LB},k}^{\star}\}_{k=r_{\text{LB},g}}^{r_{\text{LB},1}}\right).
\end{align}
Thus, by \eqref{lem:proof1-power} and \eqref{eq:proof1-inequality2}, we can show that  \eqref{eq:proof1-norminequality} holds.

Therefore, we complete the proof of Lemma~\ref{Lem:special-w}.

\section*{Appendix B: Proof of Theorem 1}\label{app:theorem 1}
First, we show $  P_{\text{QB}}^{\star}\leq  P_{\text{LB}}^{\star}$.
By Lemma~\ref{Lem:special-w}, we know that there exists an optimal solution of Problem~\ref{P1} for the layer-based scheme $\{\mathbf{w}_{\text{LB},l}^{\star}\}_{l=1}^{r_{\text{LB},1}}$ which satisfies \eqref{eq:proof1-property2} and \eqref{eq:proof1-norminequality}.  Based on $\{\mathbf{w}_{\text{LB},l}^{\star}\}_{l=1}^{r_{\text{LB},1}}$, we construct a feasible solution of Problem~\ref{P1} with $i=\text{QB}$ which achieves the same total transmission power $ P_{\text{LB}}^{\star}$ as $\{\mathbf{w}_{\text{LB},l}^{\star}\}_{l=1}^{r_{\text{LB},1}}$. Define
\begin{align}\label{eq:proof1-define1}
\mathbf{w}_{\text{QB},l}^{\dag}\coloneqq \mathbf{w}_{\text{LB},r_{\text{LB},G+1-l}}^{\star}\left(\frac{\sum_{j=r_{\text{LB},G+2-l}+1}^{r_{\text{LB},G+1-l}}\norm{\mathbf{w}_{\text{LB},j}^{\star}}^2}{\lVert\mathbf{w}_{\text{LB},r_{\text{LB},G+1-l}}^{\star}\rVert^2}\right)^{\frac{1}{2}}, ~l\in\Upsilon_{\text{QB},\text{max}}.
\end{align}
Thus, we have
\begin{align}\label{eq:proof1-ef-P2-sinr}
&|\mathbf{h}_u^H\mathbf{w}_{\text{QB},l}^{\dag}|^2=|\mathbf{h}_u^H\mathbf{w}_{\text{LB},r_{\text{LB},G+1-l}}^{\star}|^2\frac{\sum_{j=r_{\text{LB},G+2-l}+1}^{r_{\text{LB},G+1-l}}\norm{\mathbf{w}_{\text{LB},j}^{\star}}^2}{\lVert\mathbf{w}_{\text{LB},r_{\text{LB},G+1-l}}^{\star}\rVert^2}\overset{(a)}\geq|\mathbf{h}_u^H\mathbf{w}_{\text{LB},r_{\text{LB},G+1-l}}^{\star}|^2\frac{\Gamma_{\text{QB},l}}{\Gamma_{\text{LB},r_{\text{LB},G+1-l}}}\nonumber\\
&\overset{(b)}\geq\Gamma_{\text{QB},l}\left(\sum_{k=r_{\text{LB},G+1-l}+1}^{r_{\text{LB},1}}|\mathbf{h}_u^H\mathbf{w}_{\text{LB},k}^{\star}|^2+\sigma_u^2\right)=\Gamma_{\text{QB},l}\left(\sum_{j=l+1}^{G}\sum_{k=r_{\text{LB},G+2-j}+1}^{r_{\text{LB},G+1-j}}|\mathbf{h}_u^H\mathbf{\bar{w}}_{\text{LB},k}^{\star}|^2\norm{\mathbf{w}_{\text{LB},k}^{\star}}^2+\sigma_u^2\right)\nonumber\\
&\overset{(c)}=\Gamma_{\text{QB},l}\left(\sum_{j=l+1}^{G}\sum_{k=r_{\text{LB},G+2-j}+1}^{r_{\text{LB},G+1-j}}|\mathbf{h}_u^H\mathbf{\bar{w}}_{\text{LB},r_{\text{LB},G+1-j}}^{\star}|^2\norm{\mathbf{w}_{\text{LB},k}^{\star}}^2+\sigma_u^2\right)\nonumber\\
&=\Gamma_{\text{QB},l}\left(\sum_{k=l+1}^{G}|\mathbf{h}_u^H\mathbf{w}_{\text{QB},k}^{\dag}|^2+\sigma_u^2\right),~u\in\mathcal{U}_{g},l\in\Upsilon_{\text{QB},g},g\in\mathcal{G},
\end{align}
where $(a)$ is due to \eqref{eq:proof1-norminequality}, $(b)$ is due to \eqref{constr:P1-sinr}, and $(c)$ is due to \eqref{eq:proof1-property2}.
By comparing \eqref{eq:proof1-ef-P2-sinr} with \eqref{constr:P1-sinr}, we can conclude that $\{\mathbf{w}_{\text{QB},l}^{\dag}\}_{l=1}^{G}$ is a feasible solution of Problem~\ref{P1} with $i=\text{QB}$.
In addition, by \eqref{eq:proof1-define1}, we have
 \begin{align}\label{eq:proof1-optvalue}
 \sum_{l=1}^G\lVert{\mathbf{w}_{\text{QB},l}^{\dag}}\rVert^2&=\sum_{l=1}^G\norm{\mathbf{w}_{\text{LB},r_{\text{LB},G+1-l}}^{\star}}^2\frac{\sum_{j=r_{\text{LB},G+2-l}+1}^{r_{\text{LB},G+1-l}}\norm{\mathbf{w}_{\text{LB},j}^{\star}}^2}{\lVert\mathbf{w}_{\text{LB},r_{\text{LB},G+1-l}}^{\star}\rVert^2}=\sum_{l=1}^{r_{\text{LB},1}}\norm{\mathbf{w}_{\text{LB},l}^{\star}}^2= P_{\text{LB}}^{\star}.
 \end{align}
 Thus, we know that the feasible solution $\{\mathbf{w}_{\text{QB},l}^{\dag}\}_{l=1}^{G}$ achieves  total transmission power $ P_{\text{LB}}^{\star}$. Therefore, the optimal value $  P_{\text{QB}}^{\star}$ of Problem~\ref{P1} with $i=\text{QB}$  satisfies $  P_{\text{QB}}^{\star}\leq P_{\text{LB}}^{\star}$.

Next, we show $ P_{\text{LB}}^{\star}\leq  P_{\text{QB}}^{\star}$. Based on an optimal solution of Problem~\ref{P1} with $i=\text{QB}$ denoted by $\{\mathbf{w}_{\text{QB},l}^{\star}\}_{l=1}^{G}$, we construct a feasible solution of Problem~\ref{P1} which achieves the same total transmission power $ P_{\text{QB}}^{\star}$ as $\{\mathbf{w}_{\text{QB},l}^{\star}\}_{l=1}^{G}$. Define
\begin{align}\label{eq:proof1-wl}
\mathbf{w}_{\text{LB},l}^{\dag}\coloneqq\mathbf{w}_{\text{QB},r_{\text{QB},g}}^{\star}\left(\frac{\Gamma_{\text{LB},l}2^{\sum_{j=l+1}^{r_{\text{LB},g}}R_{\text{LB},j}}}{\Gamma_{\text{QB},r_{\text{QB},g}}}\right)^{\frac{1}{2}},~l\in\{r_{\text{LB},g+1}+1,\ldots,r_{\text{LB},g}\},g\in\mathcal{G}.
\end{align}
Thus, we have
\begin{align}\label{eq:proof1-ef-P1-sinr}
&|\mathbf{h}_u^H\mathbf{w}_{\text{LB},l}^{\dag}|^2=|\mathbf{h}_u^H\mathbf{w}_{\text{QB},r_{\text{QB},g}}^{\star}|^2\frac{\Gamma_{\text{LB},l}2^{\sum_{j=l+1}^{r_{\text{LB},g}}R_{\text{LB},j}}}{\Gamma_{\text{QB},r_{\text{QB},g}}}\nonumber\\
&=\Gamma_{\text{LB},l}\left(\sum_{n=l+1}^{r_{\text{LB},g}}\frac{|\mathbf{h}_u^H\mathbf{w}_{\text{QB},r_{\text{QB},g}}^{\star}|^2 2^{\sum_{j=n+1}^{r_{\text{LB},g}}R_{\text{LB},j}}\Gamma_{\text{LB},n}}{\Gamma_{\text{QB},r_{\text{QB},g}}}+\frac{|\mathbf{h}_u^H\mathbf{w}_{\text{QB},r_{\text{QB},g}}^{\star}|^2}{\Gamma_{\text{QB},r_{\text{QB},g}}}\right)\nonumber\\
&=\Gamma_{\text{LB},l}\left(\sum_{n=l+1}^{r_{\text{LB},g}}|\mathbf{h}_u^H\mathbf{w}_{\text{LB},n}^{\dag}|^2+\frac{|\mathbf{h}_u^H\mathbf{w}_{\text{QB},r_{\text{QB},g}}^{\star}|^2}{\Gamma_{\text{QB},r_{\text{QB},g}}}\right)\overset{(d)}\geq\Gamma_{\text{LB},l}\left(\sum_{n=l+1}^{r_{\text{LB},g}}|\mathbf{h}_u^H\mathbf{w}_{\text{LB},n}^{\dag}|^2+\sum_{n=r_{\text{QB},g}+1}^{G}|\mathbf{h}_u^H\mathbf{w}_{\text{QB},n}^{\star}|^2+\sigma_u^2\right)\nonumber\\
&=\Gamma_{\text{LB},l}\left(\sum_{n=l+1}^{r_{\text{LB},g}}|\mathbf{h}_u^H\mathbf{w}_{\text{LB},n}^{\dag}|^2+\sum_{n=r_{\text{QB},g}+1}^{G}\sum_{j=r_{G-n+2}+1}^{r_{G-n+1}}|\mathbf{h}_u^H\mathbf{w}_{\text{LB},j}^{\dag}|^2+\sigma_u^2\right)\nonumber\\
&=\Gamma_{\text{LB},l}\left(\sum_{n=l+1}^{r_{\text{LB},1}}|\mathbf{h}_u^H\mathbf{w}_{\text{LB},n}^{\dag}|^2+\sigma_u^2\right),~u\in\mathcal{U}_g,\in\{r_{\text{LB},g+1}+1,\ldots,r_{\text{LB},g}\},g\in\mathcal{G},
\end{align}
where $(d)$ is due to \eqref{constr:P1-sinr}. By comparing \eqref{eq:proof1-ef-P1-sinr} with \eqref{constr:P1-sinr}, we can conclude that $\{\mathbf{w}_{\text{LB},l}^{\dag}\}_{l=1}^{r_{\text{LB},1}}$  is a feasible solution of Problem~\ref{P1}.
In addition, by \eqref{eq:proof1-wl}, we have
\begin{align}
&\sum_{l=1}^{r_{\text{LB},1}}\lVert\mathbf{w}_{\text{LB},l}^{\dag}\rVert^2=\sum_{g=1}^{G}\left(\lVert\mathbf{w}_{\text{QB},r_{\text{QB},g}}^{\star}\rVert^2\sum_{l=r_{\text{LB},g+1}+1}^{r_{\text{LB},g}}\frac{\Gamma_{\text{LB},l}2^{\sum_{j=l+1}^{r_{\text{LB},g}}R_{\text{LB},j}}}{\Gamma_{\text{QB},r_{\text{QB},g}}}\right)=\sum_{l=1}^{G}\lVert\mathbf{w}_{\text{QB},l}^{\star}\rVert^2=  P_{\text{QB}}^{\star}.
\end{align}
Thus, we know that the feasible solution $\{\mathbf{w}_{\text{LB},l}^{\dag}\}_{l=1}^{r_{\text{LB},1}}$ achieves total transmission power $ P_{\text{QB}}^{\star}$. Therefore, the optimal value $ P_{\text{LB}}^{\star}$ of Problem~\ref{P1} satisfies
$ P_{\text{LB}}^{\star}\leq  P_{\text{QB}}^{\star}$.
Since $  P_{\text{QB}}^{\star}\leq  P_{\text{LB}}^{\star}$ and $ P_{\text{LB}}^{\star}\leq  P_{\text{QB}}^{\star}$, we have $ P_{\text{LB}}^{\star}=  P_{\text{QB}}^{\star}$. Therefore, we complete the proof of Theorem~\ref{thm:P12}.

\section*{Appendix C: Proof of Theorem 2}\label{app:theorem 2}
First, we show $ P_{\text{QB},\text{SDR}}^{\star}\leq P_{\text{LB},\text{SDR}}^{\star}$. Based on an optimal solution of Problem~\ref{SDR1} with $i=\text{LB}$ denoted as $\{\mathbf{X}_l^{\star}\}_{l=1}^{r_{\text{LB},1}}$, we construct a feasible solution of Problem~\ref{SDR1} with $i=\text{QB}$ which achieves objective value $P_{\text{LB},\text{SDR}}^{\star}$. Define
\begin{align}\label{eq:proof2-define1}
\mathbf{X}_{\text{QB},l}^{\dag}\coloneqq\sum_{j=r_{\text{LB},G+2-l}+1}^{r_{\text{LB},G+1-l}}\mathbf{X}_{\text{LB},j}^{\star},~l\in\Gamma_{\text{QB},\text{max}}.
\end{align}
Thus, we have
\begin{align}\label{eq:proof2-ef-SDR2-sinr}
&\text{tr}(\mathbf{C}_u\mathbf{X}_{\text{QB},l}^{\dag})=\sum_{j=r_{\text{LB},G+2-l}+1}^{r_{\text{LB},G+1-l}}\text{tr}(\mathbf{C}_u\mathbf{X}_{\text{LB},j}^{\star})\overset{(a)}\geq\sum_{j=r_{\text{LB},G+2-l}+1}^{r_{\text{LB},G+1-l}}\Gamma_{\text{LB},j} \left(\sum_{k=j+1}^{r_{\text{LB},1}}\text{tr}(\mathbf{C}_u\mathbf{X}_{\text{LB},k}^{\star})+\sigma_u^2\right)\nonumber\\
&\overset{(b)}\geq\sum_{j=r_{\text{LB},G+2-l}+1}^{r_{\text{LB},G+1-l}}\Gamma_{\text{LB},j} 2^{\sum_{n=j+1}^{r_{\text{LB},G+1-l}}R_{\text{LB},n}}\left(\sum_{k=r_{\text{LB},G+1-l}+1}^{r_{\text{LB},1}}\text{tr}(\mathbf{C}_u\mathbf{X}_{\text{LB},k}^{\star})+\sigma_u^2\right)\nonumber\\
&=\Gamma_{\text{QB},l}\left(\sum_{k=l+1}^{G}\text{tr}(\mathbf{C}_u\mathbf{X}_{\text{QB},k}^{\dag})+\sigma_u^2\right),~u\in\mathcal{U}_g,l\in\Gamma_{\text{QB},g},g\in\mathcal{G},
\end{align}
where $(a)$ and $(b)$ are due to \eqref{constr:EP1-sinr}. By comparing \eqref{eq:proof2-ef-SDR2-sinr} with \eqref{constr:EP1-sinr}, we can conclude that $\{\mathbf{X}_{\text{QB},l}^{\dag}\}_{l=1}^{G}$ is a feasible solution of Problem~\ref{SDR1} with $i=\text{QB}$. In addition, by \eqref{eq:proof2-define1}, we have
\begin{align}\label{eq:proof2-power1}
\sum_{l=1}^{G}\text{tr}(\mathbf{X}_{\text{QB},l}^{\dag})&=\sum_{l=1}^{G}\sum_{j=r_{\text{LB},G+2-l}+1}^{r_{\text{LB},G+1-l}}\text{tr}(\mathbf{X}_{\text{LB},j}^{\star})=\sum_{l=1}^{r_{\text{LB},1}}\text{tr}(\mathbf{X}_{\text{LB},l}^{\star})=P_{\text{LB}}^{\star}.
\end{align}
Thus, we know that the feasible solution $\{\mathbf{X}_{\text{QB},l}^{\dag}\}_{l=1}^{G}$ achieves objective value $P_{\text{LB},\text{SDR}}^{\star}$. Therefore, the optimal value  $ P_{\text{QB},\text{SDR}}^{\star}$ of Problem~\ref{SDR1} with $i=\text{QB}$ satisfies $ P_{\text{QB},\text{SDR}}^{\star}\leq P_{\text{LB}}^{\star}$.

Next, we show $P_{\text{LB},\text{SDR}}^{\star}\leq P_{\text{QB},\text{SDR}}^{\star}$. Based on an optimal solution of Problem~\ref{SDR1} with $i=\text{QB}$ denoted by $\{\mathbf{X}_{\text{QB},l}^{\star}\}_{l=1}^{G}$, we construct a feasible solution of Problem~\ref{SDR1} which achieves objective value $ P_{\text{QB},\text{SDR}}^{\star}$. Define
\begin{align}\label{eq:proof2-define2}
\mathbf{X}_{\text{LB},l}^{\dag}\coloneqq\mathbf{X}_{\text{QB},r_{\text{QB},g}}^{\star}\frac{\Gamma_{\text{LB},l}2^{\sum_{j=l+1}^{r_{\text{LB},g}}R_{\text{LB},j}}}{\Gamma_{\text{QB},r_{\text{QB},g}}},~l\in\{r_{\text{LB},g+1}+1,\ldots,r_{\text{LB},g}\},g\in\mathcal{G}.
\end{align}
Thus, we have
\begin{align}\label{eq:proof2-ef-SDR1-sinr}
&\text{tr}(\mathbf{C}_u\mathbf{X}_{\text{LB},l}^{\dag})\overset{(c)}=\text{tr}(\mathbf{C}_u\mathbf{X}_{\text{QB},r_{\text{QB},g}}^{\star})\frac{\Gamma_{\text{LB},r_{\text{QB},g}}2^{\sum_{j=l+1}^{r_{\text{LB},g}}R_{\text{LB},j}}}{\Gamma_{\text{QB},l}}=\Gamma_{\text{LB},l}\left(\sum_{n=l+1}^{r_{\text{LB},g}}\text{tr}(\mathbf{C}_u\mathbf{X}_{\text{LB},n}^{\dag})+\frac{\text{tr}(\mathbf{C}_u\mathbf{X}_{\text{QB},r_{\text{QB},g}}^{\star})}{\Gamma_{\text{QB},r_{\text{QB},g}}}\right)\nonumber\\
&\overset{(d)}\geq\Gamma_{\text{LB},l}\left(\sum_{n=l+1}^{r_{\text{LB},g}}\text{tr}(\mathbf{C}_u\mathbf{X}_{\text{LB},n}^{\dag})+\sum_{n=r_{\text{QB},g}+1}^{G}\text{tr}(\mathbf{C}_u\mathbf{X}_{\text{QB},n}^{\star})+\sigma_u^2\right)\nonumber\\
&\overset{(e)}=\Gamma_{\text{LB},l}\left(\sum_{n=l+1}^{r_{\text{LB},1}}\text{tr}(\mathbf{C}_u\mathbf{X}_{\text{LB},n}^{\dag})+\sigma_u^2\right),~u\in\mathcal{U}_g,l\in\{r_{\text{LB},g+1}+1,\ldots,r_{\text{LB},g}\},g\in\mathcal{G},
\end{align}
where $(c)$ is due to \eqref{eq:proof2-define2}, $(d)$ is due to \eqref{constr:EP1-sinr} and $(e)$ is due to \eqref{eq:proof2-define2}. By comparing \eqref{eq:proof2-ef-SDR1-sinr} with \eqref{constr:EP1-sinr}, we can conclude that $\{\mathbf{X}_{\text{LB},l}^{\dag}\}_{l=1}^{r_{\text{LB},1}}$  is a feasible solution of Problem~\ref{SDR1}. In addition, by \eqref{eq:proof2-define2}, we have
\begin{align}\label{eq:proof2-power2}
&\sum_{l=1}^{r_{\text{LB},1}}\text{tr}(\mathbf{X}_{\text{LB},l}^{\dag})=\sum_{g=1}^{G}\left(\text{tr}(\mathbf{X}_{\text{QB},l}^{\star})\sum_{l=r_{\text{LB},g+1}+1}^{r_{\text{LB},g}}\frac{\Gamma_{\text{LB},l}2^{\sum_{j=l+1}^{r_{\text{LB},g}}R_{\text{LB},j}}}{\Gamma_{\text{QB},r_{\text{QB},g}}}\right)=\sum_{l=1}^{G}\text{tr}(\mathbf{X}_{\text{QB},l}^{\star})=P_{\text{QB},\text{SDR}}^{\star}.
\end{align}
Thus, we know that the feasible solution $\{\mathbf{X}_{\text{LB},l}^{\dag}\}_{l=1}^{r_{\text{LB},1}}$ achieves  objective value $ P_{\text{QB},\text{SDR}}^{\star}$. Therefore, the optimal value $P_{\text{LB},\text{SDR}}^{\star}$ of Problem~\ref{SDR1} satisfies $P_{\text{LB},\text{SDR}}^{\star}\leq P_{\text{QB},\text{SDR}}^{\star}$.
Since $ P_{\text{QB},\text{SDR}}^{\star}\leq P_{\text{LB},\text{SDR}}^{\star}$ and $P_{\text{LB},\text{SDR}}^{\star}\leq P_{\text{QB},\text{SDR}}^{\star}$, we have $P_{\text{LB},\text{SDR}}^{\star}= P_{\text{QB},\text{SDR}}^{\star}$. Therefore, we complete the proof of Theorem~\ref{thm:SDR12}.

\section*{Appendix D: Proof of Lemma 2}\label{app:lemma 2}
We  rewrite the optimal value of Problem~\ref{SDR1} with $i=\text{LB}$ for given $\mathbf{r}_{\text{LB}}$ and $\mathbf{H}$ as $P_{\text{LB},\text{SDR}}^{\star}(\mathbf{r}_{\text{LB}},\mathbf{H})$.
Let
$ P_{\text{LB},\text{SDR},\text{max}}^{\star}(\mathbf{r}_{\text{LB}})\coloneqq\max\limits_{\mathbf{H}\in\mathbf{\mathcal{H}}}P_{\text{LB},\text{SDR}}^{\star}(\mathbf{r}_{\text{LB}},\mathbf{H})
$.
By Theorem~\ref{thm:SDR12}, we know that  $P_{\text{LB},\text{SDR}}^{\star}(\mathbf{r}_{\text{LB}},\mathbf{H})= P_{\text{QB},\text{SDR}}^{\star}(\mathbf{r}_{\text{LB}},\mathbf{H})$. Thus, $ P_{\text{LB},\text{SDR},\text{max}}^{\star}(\mathbf{r}_{\text{LB}})= P_{\text{QB},\text{SDR},\text{max}}^{\star}(\mathbf{r}_{\text{LB}})$. To prove Lemma~\ref{lemma2}, it remains to show that if  $\mathbf{r}_{\text{LB},1}\succcurlyeq\mathbf{r}_{\text{LB},2}$, then $ P_{\text{LB},\text{SDR},\text{max}}^{\star}(\mathbf{r}_{\text{LB},1})\geq P_{\text{LB},\text{SDR},\text{max}}^{\star}(\mathbf{r}_{\text{LB},2})$. In the following, we consider Problem~\ref{SDR1} with $i=\text{LB}$.
W.l.o.g., we consider $r_{\text{LB},2,g^{\star}}=r_{\text{LB},1,g^{\star}}-1$, and $r_{\text{LB},2,g}=r_{\text{LB},1,g}$ for all $g\in\{g|g\neq g^{\star}, g\in\mathcal{G}\}$, where $r_{\text{LB},i,g}$ denotes the $g$-th element of $\mathbf{r}_{\text{LB},i}$ for $i\in\{1,2\}$.  Then, we have $L_{\text{LB},\text{max}}(\mathbf{r}_{\text{LB},2})\leq L_{\text{LB},\text{max}}(\mathbf{r}_{\text{LB},1})$, where  $L_{\text{LB},\text{max}}(\mathbf{r}_{\text{LB}})\coloneqq\max\{r_{\text{LB},1},\ldots,r_{\text{LB},g}\}$.
We consider the following two cases. Case (\textrm{i}): $L_{\text{LB},\text{max}}(\mathbf{r}_{\text{LB},2})=L_{\text{LB},\text{max}}(\mathbf{r}_{\text{LB},1})$. In this case, the objective functions of Problem~\ref{SDR1} with  $(\mathbf{r}_{\text{LB},1},\mathbf{H})$  and that of Problem~\ref{SDR1}   with $(\mathbf{r}_{\text{LB},2},\mathbf{H})$  are the same. Besides, the constraint set of Problem~\ref{SDR1} with   $(\mathbf{r}_{\text{LB},1},\mathbf{H})$ contains that of Problem~\ref{SDR1} with $(\mathbf{r}_{\text{LB},2} ,\mathbf{H})$. Thus, we have
$P_{\text{LB},\text{SDR}}^{\star}(\mathbf{r}_{\text{LB},2},\mathbf{H})\leq P_{\text{LB},\text{SDR}}^{\star}(\mathbf{r}_{\text{LB},1},\mathbf{H})$ for all $\mathbf{H}\in\mathcal{H}$. Then, we have $P_{\text{LB},\text{SDR},\text{max}}^{\star}(\mathbf{r}_{\text{LB},2})\leq P_{\text{LB},\text{SDR},\text{max}}^{\star}(\mathbf{r}_{\text{LB},1})$.
Case (\textrm{ii}): $L_{\text{LB},\text{max}}(\mathbf{r}_{\text{LB},2})<L_{\text{LB},\text{max}}(\mathbf{r}_{\text{LB},1})$. In this case, we focus on the SINR constraint for decoding layer $L_{\text{LB},\text{max}}(\mathbf{r}_{\text{LB},1})$ successfully in \eqref{constr:EP1-sinr} of Problem~\ref{SDR1} with  $(\mathbf{r}_{\text{LB},1},\mathbf{H})$. By relaxing $\Gamma_{\text{LB},L_{\text{LB},\text{max}}(\mathbf{r}_{\text{LB},1})}$ to zero, the obtained relaxed problem is equivalent to  Problem~\ref{SDR1}  with $(\mathbf{r}_{\text{LB},2},\mathbf{H})$. Thus, we have $P_{\text{LB},\text{SDR}}^{\star}(\mathbf{r}_{\text{LB},2},\mathbf{H})\leq P_{\text{LB},\text{SDR}}^{\star}(\mathbf{r}_{\text{LB},1},\mathbf{H})$  for all $\mathbf{H}\in\mathcal{H}$.
Then, we have $ P_{\text{LB},\text{SDR},\text{max}}^{\star}(\mathbf{r}_{\text{LB},1})\geq P_{\text{LB},\text{SDR},\text{max}}^{\star}(\mathbf{r}_{\text{LB},2})$.
Therefore, we complete the proof.

\end{document}